\documentclass[aps,prd,showpacs,nofootinbib,amssymb]{revtex4}
\usepackage{epsfig}
\newcommand{\be}{\begin{equation}}
\newcommand{\ee}{\end{equation}}
\newcommand{\ba}{\begin{eqnarray}}
\newcommand{\ea}{\end{eqnarray}}
\newcommand{\Mn}{M_{\rm N}}
\newcommand{\pE}{p_{\rm E}}
\newcommand{\sigmaPiN}{\sigma_{\pi\rm N}}
\newcommand{\trF}{{\rm tr}_{\rm F}}
\newcommand{\TL}[5]{#1 & #2 & #3 & #4 & #5 \\}
\newcommand{\TLtitle}[1]{\hspace{-0.2cm}#1 & & & & \\}
\newcommand{\TLempty}{& & & & \\ }
\newcommand{\fslash}[1] {{\not\! #1\,}}
\newcommand{\di}{\!{\rm d}}
\newcommand{\la}{\langle}
\newcommand{\ra}{\rangle}

\begin{document}
\title{\boldmath 
	The sigma-term form factor of the nucleon in the large-$N_c$ limit}
\author{P.~Schweitzer}
\affiliation{Dipartimento di Fisica Nucleare e Teorica, 
  	Universit\`a degli Studi di Pavia, Pavia, Italy\\
	and Institut f\"ur Theoretische Physik II,
 	Ruhr-Universit\"at Bochum,  D-44780 Bochum, Germany}
\date{October 2003}
\begin{abstract}
The scalar isoscalar form factor $\sigma(t)$ of the nucleon is calculated 
in the limit of a large number of colours in the framework of the chiral 
quark-soliton model.
 The calculation is free of adjustable parameters and based on an approximation
justified by the small packing fraction of instantons in the QCD vacuum model,
from which the chiral quark-soliton model was derived. 
The result for $\sigma(t)$ reproduces all features of the form factor observed 
in previous exact numerical calculations in the chiral quark-soliton model
and in chiral perturbation theory, and agrees well with the available 
phenomenological information. 
The Feynman-Hellmann theorem is used to study the pion mass dependence of 
the nucleon mass and a good agreement with lattice QCD results is observed.
The use of the present method to extrapolate lattice data to the chiral limit 
is discussed.
\end{abstract} 

\maketitle

\section{Introduction}

The pion-nucleon sigma-term $\sigmaPiN$ is of fundamental importance 
for understanding chiral symmetry breaking effects in the nucleon 
\cite{review-early,review-recent}.
It can be inferred from pion-nucleon scattering data \cite{low-energy-theorem}
and the value is found $\sigmaPiN\simeq(50-70)\,{\rm MeV}$ 
\cite{Koch:pu,Gasser:1990ce,Kaufmann:dd,Olsson:1999jt,Pavan:2001wz,Olsson:pi}.
The pion-nucleon sigma-term has wide phenomenological impact in many fields.
E.g., it is related to partial restoration of chiral symmetry in the nuclear 
medium \cite{Cohen:1991nk,Birse:1998iu} and to central nuclear forces 
\cite{Maekawa:2000sb}. It is an important ingredient in the mass decomposition
of the nucleon \cite{Ji:1994av} and plays an important role in the searches 
for the Higgs boson \cite{Cheng:1988cz}, dark matter and super-symmetric 
particles \cite{Bottino:1999ei,Chattopadhyay:2001va}.
Valuable insights into the $\sigmaPiN$-physics were provided from studies in 
chiral perturbation theory \cite{Gasser:1980sb,Bernard:1993nj,further-ChiPT,Borasoy:1998uu,Becher:1999he,Oh:1999yj},
lattice QCD \cite{Dong:1995ec,Leinweber:2000sa,Leinweber:2003dg}, 
and numerous chiral models, e.g.\  
\cite{Dominguez:kq,Diakonov:1988mg,Wakamatsu:1992wr,Kim:1995hu,linear-sigma,Adkins:1983hy,Blaizot:1988ge,Brise:fd,Jameson:ep,Stuckey:1996qr,Bernard:1987sg,Ivanov:1998wp,Lyubovitskij:2000sf,Dmitrasinovic:1999mf,Kondratyuk:2002fy}.

In this note the scalar isoscalar nucleon form factor $\sigma(t)$, which at 
zero momentum transfer $t$ yields $\sigmaPiN$, is studied in the chiral 
quark-soliton model ($\chi$QSM) \cite{Diakonov:yh,Diakonov:1987ty}.
The picture of baryons as chiral solitons of the pion field is justified 
in the limit of a large number of colours $N_c$ \cite{Witten:tx}. Despite 
the fact that in nature $N_c=3$ does not seem to be large the $\chi$QSM 
has proven to describe successfully numerous properties of the nucleon 
\cite{Christov:1995vm,Alkofer:1994ph,Diakonov:1996sr,forward-distr} and to 
provide valuable insights. E.g., in this model the observation was made of 
the so-called $D$-term in generalized parton distribution functions 
\cite{off-forward-distr} -- a fundamental characteristics of the nucleon 
\cite{Goeke:2001tz} which, when known, e.g., will provide information about 
the ``distribution of strong forces'' in the nucleon \cite{Polyakov:2002yz}.
Another recent highlight made on the basis of the soliton picture was the 
accurate prediction of the exotic $\Theta^+$ baryon \cite{Diakonov:1997mm}
for which by now strong experimental indication is collected \cite{Theta+}.

The $\chi$QSM was derived from the instanton model of the QCD vacuum 
\cite{Diakonov:1983hh,Diakonov:1985eg} and incorporates chiral symmetry 
breaking. The field theoretical character of the model plays a crucial 
role in ensuring the theoretical consistency of the approach.
Both $\sigmaPiN$ and $\sigma(t)$ were already studied in the $\chi$QSM 
in \cite{Diakonov:1988mg,Wakamatsu:1992wr} and \cite{Kim:1995hu}, where the 
model expressions were exactly evaluated using involved numerical methods. 
Instead the calculation of $\sigma(t)$ presented here is based on an 
approximation which is consistently justified by arguments from the 
instanton vacuum model.

The approximate method not only considerably simplifies the calculation --
allowing a lucid interpretation of the results and suggesting an appealing 
explanation why $\sigmaPiN$ is so large.
In addition it has methodical advantages over the exact calculations 
\cite{Diakonov:1988mg,Wakamatsu:1992wr,Kim:1995hu}. 
For in the $\chi$QSM $\sigma(t)$ is quadratically UV-divergent and as such 
strongly sensitive to the details of regularization. The virtue of the 
approximate method is that it yields a regularization scheme independent 
result. Another advantage is that the theoretical accuracy of the results -- 
which in models often has to be concluded from the comparison to phenomenology
-- is here known in advance and under theoretical control. 
To within this accuracy results from the exact $\chi$QSM calculations 
\cite{Diakonov:1988mg,Wakamatsu:1992wr,Kim:1995hu} are reproduced, 
and a good agreement with phenomenological information and results 
from other approaches is observed without any adjustable parameters. 

Another advantage is that the approximation allows to
analytically study the properties of $\sigma(t)$ in the chiral limit 
-- which is a rare occasion in the $\chi$QSM. Finally, by exploring 
the Feynman-Hellmann theorem \cite{Feynman-Hellmann-theorem}, 
the pion mass $(m_\pi)$ dependence of the nucleon mass ($\Mn$) is studied.
An exact treatment of this issue in the $\chi$QSM would be numerically highly 
involved. The results for $\Mn(m_\pi)$ compare well to lattice QCD results. 
This observation could be used to inspire ans\"atze for the chiral 
extrapolation of lattice data.

The note is organized as follows.
In Section~\ref{Sec:sigma-general} the form factor $\sigma(t)$ is defined 
and briefly discussed. In Section~\ref{Sec:model} the model is introduced.
The form factor is computed in Section~\ref{Sec:sigma-in-model} and the 
numerical results are discussed in Section~\ref{Sec:results+discussion}. 
Section~\ref{Sec:chiral-limit} is devoted to the study of 
$\sigma(t)$ and the nucleon mass as functions of the pion mass. 
Section~\ref{Sec:conclusions} contains the conclusions.
Technical details of the calculation can be found in the 
Appendices~\ref{App:gradient} and \ref{App:sigma(t)}.

\section{The scalar isoscalar nucleon form-factor}
\label{Sec:sigma-general}

The nucleon sigma-term form factor $\sigma(t)$ is defined as the form factor 
of the double commutator of the strong interaction Hamiltonian with two axial 
isovector charges \cite{review-early}.  
Disregarding a ``double isospin violating term'' proportional to 
$(m_u-m_d)(\bar{\psi}_u\psi_u-\bar{\psi}_d\psi_d)$ and commonly 
presumed to be negligible the form factor can be expressed as
\be\label{Eq-def-sigma(t)}
	\sigma(t)\,\bar{u}({\bf p'})u({\bf p}) = m \,\la N({\bf p'})|\,
	\biggl(\bar{\psi}_u(0)\psi_u(0)+\bar{\psi}_d(0)\psi_d(0)\biggr)\,
	|N({\bf p})\ra \;\;,\;\; t = (p-p')^2 \;\;,\ee
where $m = \frac12(m_u+m_d)$. In Eq.~(\ref{Eq-def-sigma(t)}) $|N({\bf p})\ra$
is the spin-averaged state of the nucleon of momentum ${\bf p}$ normalized 
as $\la N({\bf p'})|N({\bf p})\ra = 2p^0\delta^{(3)}({\bf p}-{\bf p'})$, and 
$u({\bf p})$ is the nucleon spinor with $\bar{u}({\bf p})u({\bf p})=2\Mn$.

At zero momentum transfer the form factor yields the pion-nucleon sigma-term, 
i.e. 
\be\label{Eq-def-sigmaPiN}
	\sigmaPiN\equiv\sigma(0) = \frac{m}{2\Mn}\,\la N({\bf p})|\,
	\biggl(\bar{\psi}_u(0)\psi_u(0)+\bar{\psi}_d(0)\psi_d(0)\biggr)\,
	|N({\bf p})\ra \;\;.\ee
The pion-nucleon sigma-term $\sigmaPiN$ is normalization scale invariant.
The form factor $\sigma(t)$ describes the elastic scattering off the nucleon 
due to the exchange of an isoscalar spin-zero particle. It is not known 
experimentally except for its value at the (unphysical) Cheng-Dashen point 
$t=2m_\pi^2$.
A low energy theorem \cite{low-energy-theorem} relates the value of $\sigma(t)$
at the so-called Cheng-Dashen point $t=2m_\pi^2$ to the (isoscalar even)
pion nucleon scattering amplitude. The analysis of pion-nucleon scattering
data yields
\be\label{sigma(2m_pi^2)}
	\sigma(2m_\pi^2) = \cases{
	(64\pm 8)\,{\rm MeV} & (1982) \cite{Koch:pu}\cr
	(88\pm15)\,{\rm MeV} & (1999) \cite{Kaufmann:dd}\cr
	(71\pm 9)\,{\rm MeV} & (1999) \cite{Olsson:1999jt}\cr
	(79\pm 7)\,{\rm MeV} & (2001) \cite{Pavan:2001wz}.}\ee
The recent analyses tend to yield a larger value for $\sigma(2m_\pi^2)$ 
which can be explained by the more recent and accurate data \cite{Olsson:pi}.
The difference $\sigma(2m_\pi^2)-\sigma(0)$ has been calculated from 
a dispersion relation analysis \cite{Gasser:1990ce} 
\be\label{sigma-diff}
	\sigma(2m_\pi^2)-\sigma(0) = (15.2\pm 0.4)\;{\rm MeV} \;.
\ee
In Ref.~\cite{Becher:1999he} a similar result was obtained 
from a calculation in the chiral perturbation theory. 
From Eqs.~(\ref{sigma(2m_pi^2)},~\ref{sigma-diff}) one concludes
\be\label{sigmaPiN}
	\sigmaPiN \simeq (50-70)\,{\rm MeV} \;. 
\ee
The large value of $\sigmaPiN$ has been and still is a puzzle 
\cite{review-early,review-recent}. According to a ``standard interpretation'' 
$\sigmaPiN$ can be related to the so-called strangeness content $y\equiv$
$2\la N|\bar\psi_s\psi_s|N\ra/\la N|(\bar\psi_u\psi_u+\bar\psi_d\psi_d)|N\ra$
of the nucleon as $(1-y)\sigmaPiN=\hat\sigma$, where $\hat\sigma$ can be
determined by means of chiral perturbation theory from baryon mass 
splittings,
$\hat\sigma=(35\pm5)\,{\rm MeV}$ \cite{Gasser:1980sb,Gasser:1982ap}.
The value in Eq.~(\ref{sigmaPiN}) then implies $y\sim (0.3-0.4)$
while one would expect $y\sim 0$ on the grounds of the OZI-rule.
It is worthwhile mentioning that such a sizeable value of $y$
naturally follows from a Goldstone-boson-pair mechanism
\cite{Dominguez:kq}.

\section{The chiral quark soliton model \boldmath ($\chi$QSM)}
\label{Sec:model}

This section briefly introduces the notions required in the following. 
More complete presentations of the model can be found in, e.g., 
\cite{Diakonov:1988mg,Diakonov:yh,Diakonov:1987ty}.
The $\chi$QSM is based on the effective chiral relativistic quantum field 
theory of quarks, antiquarks and Goldstone bosons defined by the partition 
function \cite{Diakonov:1985eg,Diakonov:tw,Dhar:gh}
\ba\label{eff-theory}
&&	Z_{\rm eff} = \int\!\!{\cal D}\psi\,{\cal D}\bar{\psi}\,{\cal D}U\;
	\exp\Biggl(i\int\di^4x\;\bar{\psi}\,
	(i\fslash{\partial}-M\,U^{\gamma_5}-m)\psi\Biggr) \;\;,\\
&&	U=\exp(i\tau^a\pi^a) \;\;,\;\;\;
	U^{\gamma_5} = \exp(i\gamma_5\tau^a\pi^a) = 
	\frac12(U+U^\dag)+\frac12(U-U^\dag)\gamma_5 \;.\nonumber\ea
In Eq.~(\ref{eff-theory}) $M$ is the dynamical quark mass, which is due to 
spontaneous breakdown of chiral symmetry and in general momentum dependent.
$U=\exp(i\tau^a\pi^a)$ denotes the $SU(2)$ chiral pion field and $m$
the current quark mass, which explicitly breaks the chiral symmetry. 
In many applications $m$ can be set to zero, but for certain quantities
it is convenient or even necessary to consider finite $m$.
The effective theory (\ref{eff-theory}) contains the Wess-Zumino term  
and the four-derivative Gasser-Leutwyler terms with correct coefficients
\cite{Diakonov:1987ty}.
It has been derived from the instanton model of the QCD vacuum 
\cite{Diakonov:1983hh,Diakonov:1985eg} and is valid at low energies 
below a scale set by the inverse of the average instanton size
\be\label{scale}
	\rho_{\rm av}^{-1} \approx 600\,{\rm MeV} \;. \ee
In practical calculations it is convenient to take the momentum dependent 
quark mass constant, i.e. $M(p)\to M(0) = 350\,{\rm MeV}$. In this case 
$\rho_{\rm av}^{-1}$ is to be understood as the cutoff, at which quark 
momenta have to be cut off within some appropriate regularization scheme.

It is important to remark that $(M\rho_{\rm av})^2$ is proportional to the 
parametrically small instanton packing fraction, i.e. with $R_{\rm av}$ 
denoting the average distance between instantons in Euclidean space-time
\be\label{packing-fraction}
	(M\rho_{\rm av})^2 \propto 
	\biggl(\frac{\rho_{\rm av}}{R_{\rm av}}\biggr)^{\!4} \ll 1 \;.\ee
Numerically $\rho_{\rm av}/R_{\rm av}\sim 1/3$. 
The parameterical smallness of this quantity played an important role in the
derivation of the effective theory (\ref{eff-theory}) from the instanton
model of the QCD-vacuum \cite{Diakonov:yh,Diakonov:1983hh}.

The $\chi$QSM is an application of the effective theory (\ref{eff-theory}) 
to the description of baryons \cite{Diakonov:yh,Diakonov:1987ty}.
The large-$N_c$ limit allows to solve the path integral over pion field 
configurations in Eq.~(\ref{eff-theory}) in the saddle-point approximation.
In the leading order of the large-$N_c$ limit the pion field is static, and 
one can determine the spectrum of the one-particle Hamiltonian of the 
effective theory (\ref{eff-theory})
\be\label{Hamiltonian}
	\hat{H}|n\ra=E_n |n\ra \;,\;\;
	\hat{H}=-i\gamma^0\gamma^k\partial_k+\gamma^0MU^{\gamma_5}+\gamma^0m 
	\;. \ee
The spectrum consists of an upper and a lower Dirac continuum, distorted by 
the pion field as compared to continua of the free Dirac-Hamiltonian
\be\label{free-Hamiltonian} 
	\hat{H}_0|n_0\ra = E_{n_0}|n_0\ra \;\;,\;\;\;\; 
	\hat{H}_0 = -i\gamma^0\gamma^k\partial_k+\gamma^0 M+\gamma^0m\;, \ee
and of a discrete bound state level of energy $E_{\rm lev}$, 
if the pion field is strong enough.
By occupying the discrete level and the states of the lower continuum each 
by $N_c$ quarks in an anti-symmetric colour state, one obtains a state 
with unity baryon number. 
The soliton energy $E_{\rm sol}$ is a functional of the pion field
\be\label{soliton-energy}
	E_{\rm sol}[U] = N_c \biggl(E_{\rm lev}+
	\sum\limits_{E_n<0}(E_n-E_{n_0})\biggr)\biggr|_{\rm reg} \;. \ee 
$E_{\rm sol}[U]$ is logarithmically divergent and has to be regularized 
appropriately, which is indicated in Eq.~(\ref{soliton-energy}).
Minimization of $E_{\rm sol}[U]$ determines the self-consistent solitonic 
pion field $U_c$.  This procedure is performed for symmetry reasons in 
the so-called hedgehog ansatz 
\be\label{hedgehog}
	\pi^a({\bf x})=e^a_r\;P(r) \;,\;\; 
	U({\bf x})=\cos P(r)+i \tau^a e_r^a \sin P(r)\;,\ee
with the radial (soliton profile) function $P(r)$ and $r=|{\bf x}|$,
${\bf e}_r = {\bf x}/r$.
The nucleon mass $\Mn$ is given by $E_{\rm sol}[U_c]$.
The momentum and the spin and isospin quantum numbers of the baryon 
are described by considering zero modes of the soliton.
Corrections in the $1/N_c$-expansion can be included by considering 
time dependent pion field fluctuations around the solitonic solution.
The $\chi$QSM provides a practical realization of the
large-$N_c$ picture of the nucleon \cite{Witten:tx}.

The self-consistent profile satisfies $P_c(0)=-\pi$ and decays in the chiral 
limit as $1/r^2$ at large $r$.  For finite $m$ it exhibits a Yukawa tail 
$\propto\exp(-m_\pi r)/r$ with the pion mass $m_\pi$ connected to $m$ 
by the Gell-Mann--Oakes--Renner relation, see below Eq.~(\ref{GMOR}).
An excellent approximation to the self-consistent profile, which exhibits all 
those features (and is sufficient for our purposes) is given by the analytical
``arctan-profile'' 
\be\label{arctan-profile}
	P(r) = -2\,{\rm arctan}\biggl(\frac{R^2_{\rm sol}}{r^2}(1+m_\pi r)
	e^{-m_\pi r} \biggr)\;,\;\; R_{\rm sol} = M^{-1} \;. \ee
The quantity $R_{\rm sol}$ is referred to as the soliton size. 
It is related to the nucleon axial coupling constant $g_A = 1.25$ as
\cite{Diakonov:1987ty,Adkins:1983ya} (note that the order of the limits 
cannot be inversed)
\be\label{rel-gA-Rsol}
	\lim\limits_{r\to\infty}
	\biggl(\lim\limits_{m_\pi\to0} r^2 P(r)\biggr) 
	= -2R_{\rm sol}^2 = -\frac{3}{8\pi}\;\frac{g_A}{f_\pi^2} \;. \ee

The $\chi$QSM allows to evaluate without adjustable parameters nucleon 
matrix elements of QCD quark bilinear operators as
\be\label{matrix-elements}	
	\la N({\bf p}')|\bar{\psi}(0)\Gamma\psi(0)|N({\bf p})\ra
	 =  c_{\mbox{\tiny$\Gamma$}} \Mn N_c \sum\limits_{n, \rm occ}
	\int\di^3{\bf x}\; e^{i({\bf p'}-{\bf p}){\bf x}}
	\bar{\Phi}_n({\bf x})\Gamma\Phi_n({\bf x})\,\bigl|_{\rm reg}
	 + \dots\;\;\ee
where $\Gamma$ is some Dirac- and flavour-matrix, $c_{\mbox{\tiny$\Gamma$}}$ 
is a constant depending on $\Gamma$ and the spin and flavour quantum numbers 
of the nucleon state $|N\ra=|S_3,T_3\ra$, and $\Phi_n({\bf x})=\la{\bf x}|n\ra$
are the coordinate space wave-functions of the single quark states $|n\ra$ 
defined in (\ref{Hamiltonian}). 
The sum in Eq.~(\ref{matrix-elements}) goes over occupied levels $n$
(i.e. $n$ with $E_n\le E_{\rm lev}$), and vacuum subtraction is implied 
for $E_n < E_{\rm lev}$ analog to Eq.~(\ref{soliton-energy}).
The dots denote terms subleading in the $1/N_c$-expansion 
(which can be included but will not be considered in this work). 
The model expressions can contain UV-divergences which have 
to be regularized as indicated in (\ref{matrix-elements}).

If in QCD the quantity on the left hand side of Eq.~(\ref{matrix-elements}) is 
normalization scale dependent, the model result refer to a scale 
of ${\cal O}(\rho_{\rm av}^{-1})$, see Eq.~(\ref{scale}).
In the way sketched in (\ref{matrix-elements}) a large variety of static 
nucleon properties like form-factors, axial properties, etc., were computed 
(see \cite{Christov:1995vm,Alkofer:1994ph} for reviews).
In Ref.~\cite{Diakonov:1996sr} the approach was generalized to non-local 
quark bilinear operators on the left hand side of (\ref{matrix-elements}) 
which paved the way to the study of the quark and antiquark distribution 
functions \cite{Diakonov:1996sr,forward-distr} and off-forward distribution 
functions \cite{off-forward-distr}.
The model results agree typically to within $(10-30)\%$ with experimental 
data or phenomenological information.

\section{\boldmath $\sigma(t)$ in the $\chi$QSM}
\label{Sec:sigma-in-model}

In this Section first the model expression for $\sigma(t)$ is discussed and
the consistency of the approach is demonstrated. Next the UV-behaviour of 
$\sigma(t)$ is studied and the question of regularization is addressed. 
Thereby is defined and justified the approximation in which then 
$\sigma(t)$ is evaluated in the next Section~\ref{Sec:results+discussion}.

\paragraph*{Expression and consistency.}
The pion-nucleon sigma-term $\sigmaPiN$ was studied in the framework of the 
$\chi$QSM in \cite{Diakonov:1988mg,Wakamatsu:1992wr} and the scalar isoscalar
form-factor $\sigma(t)$ in \cite{Kim:1995hu}. 
In leading order of the large $N_c$ limit the model expression for the 
form-factor $\sigma(t)$ reads \cite{Kim:1995hu} (in the SU(2) flavour sector)
\be\label{sigma-t-mod}
	\sigma(t) 
	= m\;N_c\int\!\di^3{\bf x}\;j_0(\sqrt{-t}\,|{\bf x}|) 
	\sum\limits_{n,\,\rm occ}\Phi^\ast_n({\bf x})\gamma^0 \Phi_n({\bf x})
	\bigl|_{\rm reg}\;,\ee
where the Bessel function $j_0(z)=\sin z\,/\,z$.  In the large-$N_c$ limit the
nucleon mass $\Mn={\cal O}(N_c)$ while the nucleon momenta $|{\bf p}|$ and
$|{\bf p'}|={\cal O}(N_c^0)$ such that $t=-({\bf p}'-{\bf p})^2$. 
Therefore (\ref{sigma-t-mod}) is valid for $|t|\ll\Mn^2$. 
(Interestingly, in the case of electromagnetic form factors the model results
agree well with data up to $|t|\sim 1\,{\rm GeV}^2$ \cite{Christov:1995vm}.)
Eq.~(\ref{sigma-t-mod}) shows that $\sigma(t)={\cal O}(N_c)$ in agreement 
with results from large-$N_c$ chiral perturbation theory \cite{Oh:1999yj}.

In the $\chi$QSM one is in a position to derive the model expression
for $\sigmaPiN$ in three different ways:
\ba
&&	\sigmaPiN = \lim\limits_{t\to0} \sigma(t)        \;,\label{way1}\\
&&	\sigmaPiN = m \;\frac{\partial\Mn(m)}{\partial m}\;,\label{way2}\\
&&	\sigmaPiN = m \int_0^1\di x\;(e^u+e^d+e^{\bar u}+e^{\bar d})(x)\;. 
	\label{way3}\ea
The first method (\ref{way1}) consists in continuing analytically the
form factor $\sigma(t)$ to $t=0$. The second method (\ref{way2}) uses the 
Feynman-Hellmann theorem \cite{Feynman-Hellmann-theorem}. The third method 
(\ref{way3}) uses the sum rule for the first moment of the flavour singlet 
twist-3 chirally odd distribution function $e^a(x)$ 
\cite{Jaffe:1991kp}.\footnote{
	Eq.~(\ref{way3}) is correct in a formal mathematical sense 
	\cite{Jaffe:1991kp}. However, the sum rule (\ref{way3}) is 
	saturated by a $\delta$-function at $x=0$ in $e^a(x)$ which means 
	that the relation (\ref{way3}) is practically useless to extracted
	any information on $\sigmaPiN$ from possible measurements of
	$e^a(x)$ in deeply inelastic scattering experiments, 
	see \cite{Efremov:2002qh} and references therein.}
The three methods consistently yield
\be\label{sigmaPiN-mod}
	\sigmaPiN
	= m\;N_c\int\!\di^3{\bf x}\;\sum\limits_{n,\,\rm occ}
	\Phi^\ast_n({\bf x})\gamma^0 \Phi_n({\bf x})\bigl|_{\rm reg} \;.\ee
The result in Eq.~(\ref{sigmaPiN-mod}) immediately follows from the model 
expression (\ref{sigma-t-mod})  (recalling $j_0(z) \to 1$ for $z \to 0$).
Relation (\ref{way2}) was used to numerically compute $\sigmaPiN$ in 
\cite{Diakonov:1988mg}, and explicitly demonstrated to yield the expression 
in Eq.~(\ref{sigmaPiN-mod}) in \cite{Schweitzer:2003uy}. 
The sum rule (\ref{way3}) was shown to be satisfied in the $\chi$QSM 
(and to yield (\ref{sigmaPiN-mod}) for $\sigmaPiN$) in 
\cite{Schweitzer:2003uy,Wakamatsu:2003uu}.
The fact that $\sigmaPiN$ can consistently be computed in the $\chi$QSM in 
three different ways illustrates the theoretical consistency of the model.

\paragraph*{The UV-behaviour of $\sigma(t)$.}
In this paragraph the known result, cf.\  
\cite{Diakonov:1988mg,Wakamatsu:1992wr,Kim:1995hu}, will be rederived that 
the model expression for $\sigma(t)$ contains quadratic and logarithmic 
divergences, i.e.\  that it is of the form
\be\label{eval-00}
	\sigma(t) 
=a_2(t)\;\biggl(\frac{\Lambda_{\rm cut}}{M}\biggr)^{\!2}_{\!\rm reg}
+a_{\rm log}(t)\;{\rm log}\biggl(\frac{\Lambda_{\rm cut}}{M}\biggr)_{\!\rm reg}
+a_0(t)\;\biggl(\frac{\Lambda_{\rm cut}}{M}\biggr)^{\!0} 
	\;\;,\ee
where the coefficients $a_i(t)$ are UV-finite functions of $t$ and
$\Lambda_{cut}$ is an UV-cutoff. A similar study was presented in 
Ref.~\cite{Schweitzer:2003uy}, however, for the more involved case 
of the twist-3 distribution function $e^a(x)$ which is related to
$\sigmaPiN$ by means of the sum rule (\ref{way3}).

Let us separately consider 
the contributions from the discrete level and the negative continuum
\ba
	\sigma(t) = \sigma(t)_{\rm lev}+\sigma(t)_{\rm cont} \;,
	&& \sigma(t)_{\rm lev}  
	= m\;N_c\int\!\di^3{\bf x}\;j_0(\sqrt{-t}\,|{\bf x}|)\; 
	  \Phi^\ast_{\rm lev}({\bf x})\gamma^0 \Phi_{\rm lev}({\bf x})\;,
	\label{eval-01}\\
	&& \sigma(t)_{\rm cont} 
	= m\;N_c\int\!\di^3{\bf x}\;j_0(\sqrt{-t}\,|{\bf x}|) 
	\sum\limits_{E_n < 0}\Phi^\ast_n({\bf x})\gamma^0 \Phi_n({\bf x})
	\bigl|_{\rm reg}\;.
	\label{eval-02}\ea
For the discrete level the eigenvalue problem
$\hat{H}\Phi_{\rm lev}({\bf x}) = E_{\rm lev}\Phi_{\rm lev}({\bf x})$ can 
exactly be solved numerically, see e.g.\  \cite{Diakonov:1996sr}, 
and with $\Phi_{\rm lev}({\bf x})$ one obtains $\sigma(t)_{\rm lev}$.
The contribution of the discrete level is always finite and for our purposes 
it is sufficient to note
\be\label{eval-01a}
	\sigma(t)_{\rm lev} = \mbox{UV-finite.} \ee

The exact evaluation of the continuum contribution in (\ref{eval-02})
is far more involved. For that one either can place the soliton in a finite 
3-D box, discretize and make finite the spectrum of the free Hamiltonian
(\ref{free-Hamiltonian}) by imposing boundary conditions and diagonalize 
the Hamiltonian (\ref{Hamiltonian}) in the basis of the free Hamiltonian 
states (Kahana-Ripka method \cite{Kahana:be}). Alternatively one can 
rewrite the continuum contribution (\ref{eval-02}) in terms of
Green functions and evaluate those by means of phase shift methods
(see, e.g., Ref.~\cite{Baacke:1998nm}).
Both methods are numerically involved.

Here we will use an approximate method -- in Ref.~\cite{Diakonov:1996sr}
referred to as {\sl interpolation formula}  -- which consists in expanding 
the continuum contribution (\ref{eval-02}) in gradients of the $U$-field 
and retaining the leading order only.
The interpolation formula yields exact results in three limiting cases 
$|\nabla U|\ll M$, $|\nabla U|\gg M$ and $|\log U|\ll 1$. Therefore one 
can expect that it yields useful estimates also for the general case. 
It was observed that this method approximates exact calculations in the 
model with good accuracy \cite{Diakonov:1996sr,forward-distr}.

Let us rewrite the continuum contribution in Eq.~(\ref{eval-02}) 
(recalling the implicit vacuum subtraction) as 
\be\label{eval-03}
	\sigma(t)_{\rm cont} 
	= m\;N_c \int_C \frac{\di\omega}{2\pi i} \;{\rm Sp} \Biggl[
	\,j_0(\sqrt{-t}\,|\hat{\bf x}|)\, \gamma^0 \frac{1}{\omega+i\hat{H}}
	- (\hat{H}\to\hat{H}_0) \Biggr]_{\rm reg}\;, \ee
where the contour $C$ is defined as going along the real $\omega$-axis and
closed in the infinity in the upper half of the complex $\omega$-plane.
The original expression (\ref{eval-02}) is recovered by saturating the
functional trace with the complete set of eigenfunctions of respectively
$\hat{H}$ and $\hat{H}_0$ in
(\ref{Hamiltonian},~\ref{free-Hamiltonian}),\footnote{
	I.e.\ ${\rm Sp}[\dots - (\hat{H}\to\hat{H}_0)] \equiv $
	$\sum_{{\rm all}\,n}\la n|\dots|n\ra - $
	$\sum_{{\rm all}\,n_0} \la n_0|\dots|n_0\ra$.} 
performing the $\omega$-integration, and passing to the coordinate space 
representation. Expanding (\ref{eval-03}) in a series in gradients of
the $U$-field
\be\label{eval-04}
	\sigma(t)_{\rm cont} = 
	\sum_{k=0}^\infty\;\sigma(t)_{\rm cont}^{(k)} \;, \ee
where the index $k$ means that $\nabla U$ appears $k$-times 
in $\sigma(t)_{\rm cont}^{(k)}$,
we obtain (see Appendix~\ref{App:gradient})
\be\label{eval-04a}
	\sigma(t)_{\rm cont}^{(k)} =
	\cases{ \alpha\;B(t) 	& for $k = 0$, \cr
		\mbox{zero}	& for $k = 1$, \cr
		\mbox{UV-finite}& for $k \ge 2$,} \ee
with the constant $\alpha$ and the ``form factor'' $B(t)$ defined as
\ba\label{eval-05a}
&&	\alpha =  m \int\frac{\di^4p_E}{(2\pi)^4}\,\frac{8N_cM}{p_E^2+M^2}
	\biggl|_{\rm reg} \;\;,\\
\label{eval-05b}
&&	B(t) = \int\di^3{\bf x} \;j_0(\sqrt{-t}\,|{\bf x}|)\;
	\biggl(1- \frac{1}{2}{\rm tr}_{\rm F} U({\bf x})\biggr) \;.\ea
The factor $8N_cMm$ is included into the definition of $\alpha$ for 
later convenience. The Euclidean integral in the constant $\alpha$ 
contains quadratic and logarithmic divergences.
Combining the results in (\ref{eval-01a}) and (\ref{eval-04a},~\ref{eval-05a})
we see that the form-factor $\sigma(t)$ has the UV-behaviour (\ref{eval-00})
(and that $a_2(t) \propto a_{\rm log}(t) \propto B(t)$).

\paragraph*{Instanton motivated approximation.}
In non-renormalizable effective (low energy) theories the regularization 
procedure ``keeps the memory'' of the cutoff $\Lambda_{\rm cut}$.
In such effective theories -- for which (\ref{eff-theory}) is an example -- 
the cutoff has a physical meaning: It sets the scale below which the degrees 
of freedom of the effective theory may be considered as appropriate to 
describe the physical situation, and above which they may not be sufficient.
In the effective theory (\ref{eff-theory}) the cutoff 
\be\label{reg-AA}
	\Lambda_{\rm cut}={\cal O}(\rho_{\rm av}^{-1})\;.\ee
Using (\ref{reg-AA}) and the results of the previous paragraph we see that 
$\sigma(t)$ can be written as,
\be\label{reg-00}
	\sigma(t) = \alpha\;B(t)\cdot\biggl\{1 + 
	{\cal O}(M^2\rho_{\rm av}^2) \biggr\} \; \ee
with $\alpha$ and $B(t)$ as defined in Eqs.~(\ref{eval-05a},~\ref{eval-05b}).

Thus, in Eq.~(\ref{eval-00}) the UV-finite contributions are parametrically 
strongly suppressed with respect to the UV-divergent terms by the 
instanton packing fraction due to Eq.~(\ref{packing-fraction}). 
Since the $\chi$QSM was derived from the instanton 
vacuum model, it is consistent to use this argument based 
on Eqs.~(\ref{packing-fraction},~\ref{reg-AA}) in this context. 
Eq.~(\ref{reg-00}) defines the approximation in which $\sigma(t)$ 
will finally be evaluated, after regularizing the divergent constant 
$\alpha$ in Eq.~(\ref{eval-05a}).

\paragraph*{Regularization.}
There are several methods to regularize the divergent constant $\alpha$ 
in Eq.~(\ref{eval-05a}). Popular methods often used 
in exact model calculations are the Schwinger proper-time regularization 
(see, e.g., \cite{Christov:1995vm}) or the Pauli-Villars subtraction method 
(see, e.g., \cite{Schweitzer:2003uy,Kubota:1999hx}).
The result will be sensitive to the chosen regularization method and, 
especially for a power divergent quantity, to the precise value of 
the cutoff -- of which only the order of magnitude is known, 
cf.\  Eq.~(\ref{reg-AA}).

However, the practical problem of how to regularize the constant $\alpha$ 
in Eq.~(\ref{eval-05a}) can be solved elegantly and in a regularization 
independent way. 
Since the quark vacuum condensate $\la\bar\psi\psi\ra_{\rm vac}$ is given 
in the effective theory (\ref{eff-theory}) by \cite{Diakonov:1985eg}
\be\label{reg-01}
	\la\bar\psi\psi\ra_{\rm vac} \equiv
	\la{\rm vac}|\biggl(\bar\psi_u(0)\psi_u(0)+\bar\psi_d(0)\psi_d(0)
	\biggr) |{\rm vac}\ra = \int\!\frac{\di^4 \pE}{(2\pi)^4}\;
	\frac{(-8N_cM)}{\pE^2+M^2}\biggl|_{\rm reg} \; , \ee
the constant $\alpha$ in Eq.~(\ref{eval-05a}) can be expressed as
\be\label{reg-02}
	\alpha = - \;m\;\la\bar\psi\psi\ra_{\rm vac} \;. \ee
At first glance Eq.~(\ref{reg-02}) is based on the mere observation that 
the same divergent integral appears in two different model expressions.
However, as will be discussed in the next section, the relation 
(\ref{reg-02}) is not accidental from a physical point of view 
(and implies an interesting interpretation of $\sigmaPiN$).
In the next step one can use the Gell-Mann--Oakes--Renner 
relation (with $f_\pi$ denoting the pion decay constant)
\be\label{GMOR}
	m_\pi^2 f_\pi^2 = -\,m\;\la\bar\psi\psi\ra_{\rm vac}\;,\ee
which is not imposed here by hand but valid in the effective theory 
(\ref{eff-theory}) \cite{Diakonov:1985eg}.
Thanks to Eqs.~(\ref{reg-02},~\ref{GMOR}) the practical problem 
of regularizing $\sigma(t)$ is shifted to the problem of regularizing 
$\la\bar\psi\psi\ra_{\rm vac}$ in Eq.~(\ref{reg-01}) or $f_\pi$.
The latter is given in the effective theory (\ref{eff-theory}) by the 
logarithmically UV-divergent expression \cite{Diakonov:1985eg}
\be\label{reg-05}
	f_\pi^2=\int\frac{\di^4 p_E}{(2\pi)^4}\;\frac{4N_cM^2}{(p_E^2+M^2)^2}
	\biggl|_{\rm reg}\;\;.\ee
Since the precise value of the cutoff is not known but only its order of 
magnitude, see Eq.~(\ref{scale}), it is customary to adjust the cutoff(s) 
in the corresponding regularization scheme such that the experimental values
of $\la\bar\psi\psi\ra_{\rm vac}$ and $f_\pi$ in (\ref{reg-01},\ref{reg-05})
are reproduced \cite{Christov:1995vm}.
In this way free parameters (cutoff, current quark mass, etc.) 
are fixed in the vacuum- and meson-sector of the effective theory 
(\ref{eff-theory}) \cite{Christov:1995vm}. 
In this sense the $\chi$QSM -- i.e.\  the baryon sector of the 
effective theory (\ref{eff-theory}) --  yields parameter-free results. 
   (In some calculations in the $\chi$QSM the mass $M$ was understood 
   as a ``free parameter'' and allowed to vary in the range 
   $350\,{\rm MeV}\le M \le 450\,{\rm MeV}$ 
   \cite{Wakamatsu:1992wr,Kim:1995hu,Christov:1995vm}. 
   The sensitivity of the model results to these variations was typically 
   within $(10-30)\%$. Here we rely on notions from the instanton vacuum 
   model and consequently take the value $M=350\,{\rm MeV}$ which follows 
   from the instanton phenomenology \cite{Diakonov:1983hh,Diakonov:1985eg}.)

Thus, our final (regularized) result for the form-factor reads
(with $B(t)$ defined in Eq.~(\ref{eval-05b}))
\be\label{reg-06}
	\sigma(t) = m_\pi^2f_\pi^2\;B(t)
	\cdot\biggl\{1 + {\cal O}(M^2\rho_{\rm av}^2) \biggr\}, \ee
where $m_\pi$ and $f_\pi$ denote the physical pion mass and decay constant, 
respectively.
It should be stressed that the result (\ref{reg-06}) does not follow
from the interpolation formula of Ref.~\cite{Diakonov:1996sr}
(which would require to add the UV-finite discrete level contribution,
cf.\  Ref.~\cite{Schweitzer:2003uy}).
Instead (\ref{reg-06}) has to be considered as an approximate result 
for $\sigma(t)$ in the $\chi$QSM, which is justified by the parametrical 
smallness of the instanton packing fraction. 

In the following the parametrically small 
${\cal O}(M^2\rho_{\rm av}^2)$-corrections often will not be indicated.

\section{\boldmath Discussion and interpretation of the results}
\label{Sec:results+discussion}

Evaluating the final expression (\ref{reg-06}) for the form factor $\sigma(t)$ 
with the soliton profile (\ref{arctan-profile}) for $M=350\,{\rm MeV}$,
$m_\pi = 140\,{\rm MeV}$ and $f_\pi=93\,{\rm MeV}$ yields the result shown in 
Figs.~1a and 1b (see Appendix~\ref{App:sigma(t)} for detailed expressions).
It should be noted that the error due to using the profile
(\ref{arctan-profile}), instead of the self-consistent profile 
which truly minimizes the soliton energy (\ref{soliton-energy}), 
is far smaller than the theoretical accuracy in Eq.~(\ref{reg-06}).

Apart from the values of $\sigma(t)$ at the Cheng-Dashen point $t=2m_\pi^2$
and at $t=0$, and their difference $\Delta_\sigma=\sigma(2m_\pi^2)-\sigma(0)$,
there is another phenomenologically interesting quantity -- namely 
the scalar mean square radius related to the slope of the form factor 
$\sigma(t)$ at $t=0$ as
\be\label{discuss-06}
	\sigma(t) = \sigmaPiN \biggl(1+\frac{1}{6}\;\la r_S^2 \ra\;t+
	{\cal O}(t^2)\biggr) \;.\ee
We obtain the results
\ba
     \sigmaPiN		&=&  67.9\,{\rm MeV} \label{fin:sigmaPiN},	\\
     \sigma(2m_\pi^2)	&=&  82.6\,{\rm MeV} \label{fin:sigma-at-CDP},	\\
     \Delta_\sigma	&=&  14.7\,{\rm MeV} \label{fin:sigma-diff},	\\
     \la r_S^2 \ra	&=&  1.00\,{\rm fm^2}\label{fin:radius}.	\ea
A useful parameterization of the form factor for negative $t$ is 
given by (with $\sigmaPiN$ from Eq.~(\ref{fin:sigmaPiN}))
\be\label{dipole-fit}
	\sigma(t) \simeq \frac{\sigmaPiN}{(1-t/M_S^2)^2}\;,\;\;\;
	M_S^2 \simeq 0.55\,{\rm GeV}^2\;.
\ee
The dipole fit (\ref{dipole-fit}) approximates $\sigma(t)$ 
to within $2\%$ for $|t|\lesssim 0.8\,{\rm GeV}^2$, cf.\  Fig.~1a.
Thus, $\sigma(t)$ decreases with increasing $|t|$ more quickly than the 
electromagnetic form factors where the corresponding dipole mass is about 
$M_{\rm em}^2\sim 0.7 {\rm GeV}^2$ in a comparable $t$-region.

The form-factor $\sigma(t)$, Eq.~(\ref{reg-06}), is not defined at and 
above the threshold $t\ge 4m_\pi^2$. In the vicinity of the threshold 
the form factor behaves as (see Appendix~\ref{App:sigma(t)})
\be\label{discuss-03}
	\sigma(t) = a_1\;\ln\Biggl(\frac{1}{1-\frac{\sqrt{t}}{2m_\pi}}\Biggr)
	          + a_2\;\;\mbox{as $t\to4m_\pi^2$ (where $t<4m_\pi^2$),} \ee
where $a_1$, $a_2$ are positive constants.
Interestingly, a similar divergent behaviour of $\sigma(t)$ for 
$t\to4m_\pi^2$ is also observed in heavy baryon chiral perturbation 
theory \cite{Bernard:1993nj,further-ChiPT,Borasoy:1998uu}.
There this feature arises as a peculiarity of the non-relativistic 
expansion and can be avoided by considering baryon chiral perturbation 
theory in manifestly Lorentz invariant form \cite{Becher:1999he}.
It is not clear whether in the $\chi$QSM this unphysical feature could also 
be cured --  possibly by a more careful analytical continuation of $\sigma(t)$ 
to $t>0$, e.g., by making use of (subtracted) dispersion relations.

%
%
	\begin{figure*}[t!]
\begin{tabular}{cc}
	\includegraphics[width=8cm,height=6.2cm]{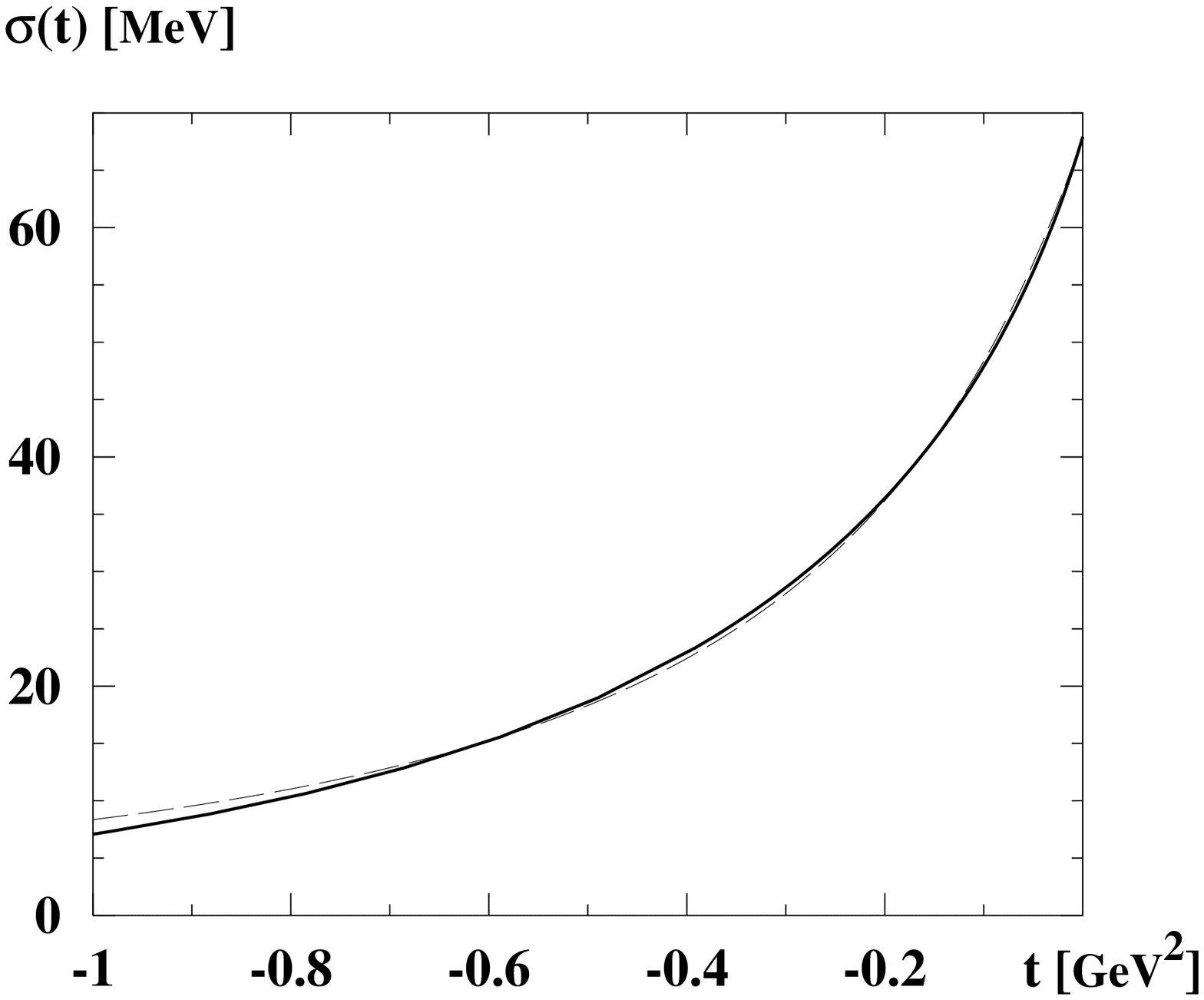} &
	\includegraphics[width=5cm,height=6.2cm]{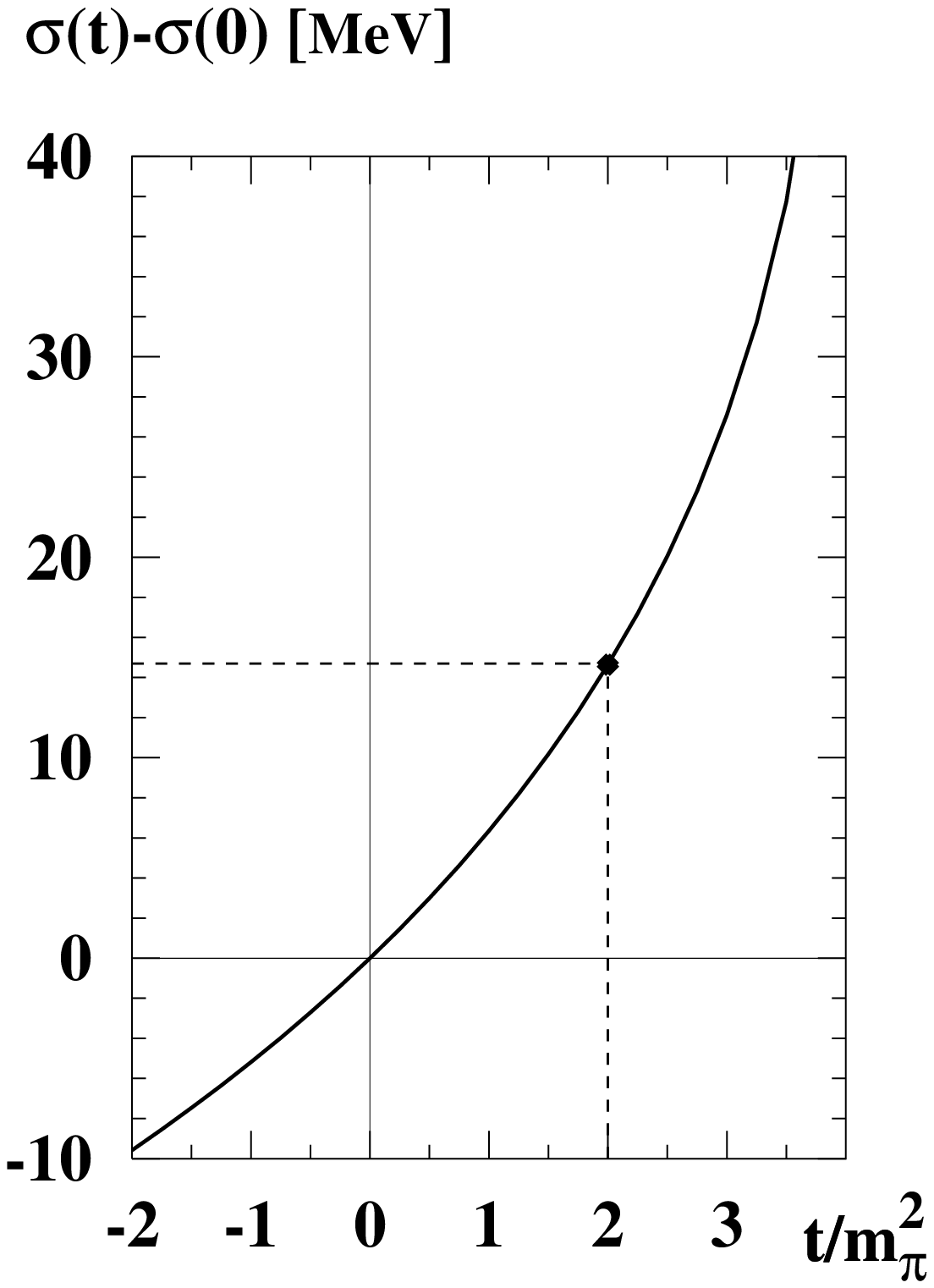}
	\cr
	{\bf a} & 
	{\bf b} 
\end{tabular}
	\caption{\footnotesize
	{\bf a.} 
	The form factor $\sigma(t)$ evaluated in the $\chi$QSM by means of 
	Eq.~(\ref{reg-06}) (solid line) and its dipole fit (\ref{dipole-fit})
	as functions of $t$.
	{\bf b}.
	The analytical continuation of the form factor $\sigma(t)$ to
	unphysical $t \ge 0$. The dot marks the value of $\sigma(t)$ 
	at the Cheng-Dashen point $t=2m_\pi^2$. At and above the threshold 
	$t\ge 4 m_\pi^2$ the form factor is not defined, see text.}
	\end{figure*}
%

\paragraph*{Accuracy of the approximation.}
Before discussing the results let us estimate the size of some of the 
contributions neglected in Eq.~(\ref{reg-06}). 
The contribution of the discrete level in (\ref{eval-01}) is 
$(\sigmaPiN)_{\rm lev} = 12\,{\rm MeV}$ \cite{Schweitzer:2003uy} 
(cf.\  next paragraph).
The contribution to the continuum part of $\sigmaPiN$ from the second order of
the gradient expansion is $(\sigmaPiN)_{\rm cont}^{(2)}\simeq-6\,{\rm MeV}$ 
(cf.\ App.~\ref{App:gradient}). 
These are corrections of ${\cal O}(15\%)$ to (\ref{fin:sigmaPiN}), i.e.\   
{\sl smaller} than the theoretical accuracy of the approximation 
(\ref{reg-06}) which is ${\cal O}(M^2\rho_{\rm av}^2)={\cal O}(30\%)$. 
This is an indication (and no more) that the approximation works.
Its theoretical justification is anyway unquestioned due to 
Eq.~(\ref{packing-fraction}).

\paragraph*{Comparison to previous calculations in the $\chi$QSM.}
In \cite{Diakonov:1988mg,Wakamatsu:1992wr} $\sigmaPiN$ was studied in SU(2)
in leading order of the large-$N_c$ expansion. In \cite{Kim:1995hu} the form
factor was studied in the SU(3) version of the model and including $1/N_c$
corrections. In \cite{Diakonov:1988mg,Wakamatsu:1992wr,Kim:1995hu} the 
proper time regularization was used. 
Our result (\ref{fin:sigmaPiN}) for $\sigmaPiN$ agrees with the numbers quoted
in \cite{Diakonov:1988mg,Wakamatsu:1992wr,Kim:1995hu} to within $30\%$.

In \cite{Kim:1995hu} it also was observed that $\sigma(t)$ is not defined for 
$t\ge 4m_\pi^2$. In the region $0<t<2m_\pi^2$ our result for the difference
$\sigma(t)-\sigma(0)$ agrees with the result of Ref.~\cite{Kim:1995hu} 
to to within $(10-20)\%$.
In the region $t<0$ the agreement of the rescaled form factor, i.e.\ 
$\sigma(t)/\sigma(0)$, is even more impressive (few percent).
It is not surprizing to observe the approximation to work 
differently in different $t$-regions.
The approximation (\ref{reg-06}) indicates the limitations of 
the numerical (finite box) method used in \cite{Kim:1995hu}. From 
Eqs.~(\ref{discuss-01},~\ref{discuss-02}) in Appendix~\ref{App:sigma(t)}
it is clear that integrals (in coordinate space) converge more and more 
slowly as $t$ approaches $t=4m_\pi^2$ (from below). 
In the finite box method, however, it is necessary that integrals converge 
quickly in order to be in the continuum (box size $\to\infty$) limit. 
Therefore, in \cite{Kim:1995hu} the region $t>2m_\pi^2$ could not be explored 
and quantitative observations of the kind (\ref{discuss-03}) were not possible.

In Ref.~\cite{Schweitzer:2003uy} the twist-3 distribution function $e^a(x)$ 
and -- by exploring the sum rule (\ref{way3}) -- also $\sigmaPiN$ were 
computed in the $\chi$QSM by means of the interpolation formula of 
Ref.~\cite{Diakonov:1996sr}.
To remind, the interpolation formula consists in estimating the continuum 
contribution essentially in the same way we did here, but to add also the
exactly evaluated discrete level contribution, which is
$(\sigmaPiN)_{\rm lev}=12\,{\rm MeV}$.
The total result $\sigmaPiN = 80\,{\rm MeV}$ of Ref.~\cite{Schweitzer:2003uy} 
agrees with the result (\ref{fin:sigmaPiN}) obtained here to within $15\,\%$.

To summarize, the instanton-motivated approximation for $\sigma(t)$,
Eq.~(\ref{reg-06}), yields results in agreement with earlier model 
calculations to within the expected accuracy.

\paragraph*{Comparison to experimental information.}
The result $\sigma(2m_\pi^2) = 83\,{\rm MeV}$ in 
Eq.~(\ref{fin:sigma-at-CDP}) agrees well with the more recently 
extracted values of $\sigma(2m_\pi^2)=(80-90)\,{\rm MeV}$
\cite{Kaufmann:dd,Olsson:1999jt,Pavan:2001wz,Olsson:pi}.
The result for $\sigma(t)-\sigmaPiN$ in the region $(-2m_\pi^2)<t<4m_\pi^2$,
cf.\  Fig.~1b, agrees with the result obtained from the dispersion relation 
analysis with chiral constraints of Ref.~\cite{Gasser:1990ce} to within 
$(2-5)\%$. 
This can be seen by comparing the difference $\Delta_\sigma$ in 
Eqs.~(\ref{sigma-diff},~\ref{fin:sigma-diff}) and the slope of $\sigma(t)$
at $t=0$, namely $\sigma'(0)\propto\sigmaPiN\,\la r_S^2\ra=72\,{\rm MeV\,fm}^2$
of \cite{Gasser:1990ce} vs.\ $68\,{\rm MeV\, fm}^2$ obtained here.
The agreement with phenomenological information can be considered as 
satisfactory.

{\def\baselinestretch{1}   
\begin{table}[t!] 
\begin{tabular}{lcccc}
\hline 
\TLempty
\TL{ }{$\sigmaPiN/{\rm MeV}$}{$\sigma(2m_\pi^2)/{\rm MeV}$}
   {$\Delta_\sigma/{\rm MeV}$}{$\la r_S^2\ra^{1/2}/{\rm fm}$}
\TL{\hspace{2cm}}{\hspace{2cm}}{\hspace{2cm}}{\hspace{2cm}}{\hspace{2cm}}
\hline
\TLempty
\TLtitle{\sl data analyses}
\TLempty
\TL{Koch (1982) \cite{Koch:pu}}                    {--}{$64\pm8$}{--}{--}
\TL{Gasser {\it et al.} (1991) \cite{Gasser:1990ce}}
	{$45\pm8$}{$60\pm8$}{$15.2\pm 0.4$}{$\sim1.3$}
\TL{Kaufmann and Hite (1999) \cite{Kaufmann:dd}}   {--}{$88\pm15$}{--}{--}
\TL{Olsson (2000) \cite{Olsson:1999jt}}		   {--}{$71\pm 9$}{--}{--}
\TL{Pavan {\it et al.} (2002) \cite{Pavan:2001wz}} {--}{$79\pm 7$}{--}{--}
\TLempty
\TLtitle{\sl theoretical approaches}
\TLempty
\TL{conventional heavy baryon $\chi$PT\cite{Borasoy:1998uu}}
	{--}{--}{$4\pm1$}{--}
\TL{manifestly Lorentz invariant $\chi$PT \cite{Becher:1999he}}{--}{--}{14}{--}
\TLempty
\TL{lattice QCD, Dong {\it et al.} (1995) \cite{Dong:1995ec}}
	{$49.7 \pm 2.6$}{--}{$6.6\pm0.6$}{$0.72 \pm 0.09$}
\TL{lattice QCD, Leinweber {\it et al.} (2000) \cite{Leinweber:2000sa}}
	{45-55}{--}{--}{--}
\TL{lattice QCD, Leinweber {\it et al.} (2003) \cite{Leinweber:2003dg}}
	{$37^{+35}_{-13}\,$-$\,73^{+15}_{-15}$}{--}{--}{--}
\TLempty
\TL{SU(2) $\chi$QSM to LO $N_c$ \cite{Diakonov:1988mg}}{54}{--}{--}{--}
\TL{SU(3) $\chi$QSM to NLO $N_c$ \cite{Kim:1995hu}}{41}{59}{18}{1.2}
\TL{linear sigma model \cite{linear-sigma}}{86-89}{--}{--}{--}
\TL{SU(2) Skyrmion \cite{Adkins:1983hy}}{49}{--}{--}{--}
\TL{SU(3) Skyrmion \cite{Blaizot:1988ge}}{60}{--}{--}{--}
\TL{chiral colour dielectric soliton \cite{Brise:fd}}{37.8}{--}{--}{1.1}
\TL{cloudy bag model \cite{Jameson:ep}}{(37-47) $\pm$ 9}{--}{--}{--}
\TL{Nambu--Jona-Lasinio \cite{Bernard:1987sg}}{50}{--}{--}{--}
\TL{confined Nambu--Jona-Lasinio \cite{Ivanov:1998wp}}{60}{--}{--}{--}
\TL{perturbative chiral quark model \cite{Lyubovitskij:2000sf}}
	{$45\pm5$}{--}{--}{--}
\TL{effective model of hadrons \cite{Kondratyuk:2002fy}}{--}{74}{--}{--}
\TLempty
\TL{This work (accuracy $\sim 30\%$)}{67.9}{82.6}{14.7}{1.00}
\TLempty
\hline 
\end{tabular}
\caption{\footnotesize 
Comparison of the results for $\sigmaPiN$, $\sigma(2m_\pi^2)$,
$\Delta_\sigma\equiv\sigma(2m_\pi^2)-\sigmaPiN$ and $\la r_S^2\ra$ from
data analyses and theoretical approaches. The list is far from complete,
only some of the more recent results are shown. For reviews on early
approaches see \cite{review-early}.}
\end{table}}

\paragraph*{Comparison to other approaches.}
Our result is consistent with the values for $\sigmaPiN$ obtained 
from lattice QCD results \cite{Dong:1995ec,Leinweber:2000sa,Leinweber:2003dg} 
(see also Section~\ref{Sec:chiral-limit}).
In Table~1 we compare our result also to calculations performed in chiral 
perturbation theory \cite{Bernard:1993nj,Borasoy:1998uu,Becher:1999he}, 
the linear sigma model \cite{linear-sigma},
Skyrme model \cite{Adkins:1983hy,Blaizot:1988ge},
colour-dielectric model \cite{Brise:fd},
cloudy bag model \cite{Jameson:ep,Stuckey:1996qr},
Nambu--Jona-Lasinio model \cite{Bernard:1987sg,Ivanov:1998wp},
perturbative chiral quark model \cite{Lyubovitskij:2000sf},
a relativistic dynamical model based on effective hadronic ($N$, $\Delta$, 
$\pi$, $\rho$ and $\sigma$) degrees of freedom \cite{Kondratyuk:2002fy}.
Also the results of the data analyses and from the $\chi$QSM-calculations
\cite{Diakonov:1988mg,Kim:1995hu} are included.

In several models it was observed that a sizable -- if not dominant -- 
contribution to $\sigmaPiN$ is due to the pion cloud (``quark sea'') 
as compared to the constituent quark core (``valence quarks'') 
\cite{Brise:fd,Jameson:ep,Stuckey:1996qr,Lyubovitskij:2000sf}.
In the $\chi$QSM the discrete level contribution corresponds to the quark core
and the continuum contribution to the pion cloud.\footnote{	
	This is to be understood in a loose sense in the ``quark model 
	language''. Strictly speaking there is no one-to-one correspondence in
	the $\chi$QSM between the discrete level and continuum contributions 
	on the one hand and the notions of ``valence'' and ``sea quarks'' 
	on the other hand. The latter are well defined, e.g., in the 
	context of parton distribution functions. The point is that both 
	discrete level and continuum contribute each to valence and sea quark 
	distributions \cite{Diakonov:1996sr,forward-distr}.}
Here we obtain an extreme picture, where $\sigmaPiN$ is purely due to 
the pion cloud. Corrections to this picture are suppressed by 
the instanton packing fraction (and are practically of order $30\%$).

\paragraph*{Skyrmion- and non-relativistic limit.}
It is possible to recover from expressions of the $\chi$QSM the results of 
the non-relativistic quark model and the Skyrme model by taking appropriate
(non-physical) limits \cite{Praszalowicz:1995vi}.

The ``Skyrmion-limit'' consists in taking $R_{\rm sol}\to\infty$. 
Since for the soliton solution $R_{\rm sol}\simeq M^{-1}$ 
\cite{Diakonov:1987ty}, the limit is to be understood as evaluating model 
quantities with, e.g., the profile in Eq.~(\ref{arctan-profile})
which allows to vary $R_{\rm sol}$ \cite{Praszalowicz:1995vi}.
In this limit the energy of the discrete level $E_{\rm lev}\to (-M)$
\cite{Diakonov:1987ty}, such that this contribution is enclosed in the 
contour of the $\omega$-integral in (\ref{eval-03}).  
With increasing $R_{\rm sol}$ the contributions 
$(\sigmaPiN)_{\rm cont}^{(k)}$ in the series (\ref{eval-04}) 
behave as $(\sigmaPiN)_{\rm cont}^{(k)} \propto R_{\rm sol}^{3-k}$.
For $(\sigmaPiN)_{\rm cont}^{(0)}$ and $(\sigmaPiN)_{\rm cont}^{(2)}$ 
this can be seen directly from the expressions in the Appendix.
For arbitrary $k$ one arrives at this conclusion using general 
scaling arguments.
Thus $(\sigmaPiN)_{\rm cont}^{(0)}$ dominates again -- 
this time, however, justified by the unphysical $R_{\rm sol}\to\infty$ limit.
The expression for $\sigmaPiN$ obtained here {\sl formally} coincides 
with the expressions in Refs.~\cite{Adkins:1983hy,Blaizot:1988ge}.
It should be noted that what is an exact result in the Skyrme model 
is here merely an approximation -- though a well justified one 
thanks to arguments from the instanton vacuum model.
The coincidence of the expressions is purely formal since the Skyrmion is a 
topological soliton. E.g., in \cite{Blaizot:1988ge} a vector-meson model 
was used to determine the $U$-field.

In the opposite limit, $R_{\rm sol}\to 0$, one recovers results from the
non-relativistic constituent quark model (formulated for arbitrary $N_c$
\cite{Karl:cz}). 
Taking $R_{\rm sol}\to 0$ in the expression (\ref{sigma-t-mod}) one obtains 
\be\label{non-rel-limit}
	\sigma(t) = \Mn G_E(t) \;,
\ee
where $G_E(t)$ (normalized to $G_E(0) = 1$) denotes the isoscalar electric 
form factor -- taken also in the non-relativistic limit. This result follows
from generalizing the discussion in Ref.~\cite{Schweitzer:2003uy} to the case 
of the form factor $\sigma(t)$. That Eq.~(\ref{non-rel-limit}) is the correct
non-relativistic relation between $\sigma(t)\propto \la N'|\bar\psi\psi|N\ra$
and $G_E(t)\propto \la N'|\psi^\dag\psi|N\ra$ follows from considering that 
in this limit the nucleon wave-function has no lower (antiquark-) component
such that $\psi^\dag\psi=\bar\psi\psi$ practically holds, and that the current 
quark mass $m$ has to be understood as a constituent quark mass equal to 
$\Mn/3$.
The non-relativistic relation $\sigmaPiN = \Mn$ strongly overestimates the 
phenomenological value of $\sigmaPiN$ in Eq.~(\ref{sigmaPiN}). 
Still, this result is theoretically consistent and, e.g., correctly implies 
a vanishing strangeness content $y$ in the nucleon \cite{Efremov:2002qh}.

\paragraph*{Interpretation of $\sigmaPiN$.}
In the effective theory (\ref{eff-theory}) the pion-nucleon sigma-term 
and the quark vacuum condensate are proportional to each other 
\be\label{propto-vac-cond}
	\sigmaPiN \propto -m\la\bar\psi\psi\ra_{\rm vac}\;,
\ee
up to parametrically small ${\cal O}(M^2\rho_{\rm av}^2)$-corrections.
From a physical point of view this observation based on Eq.~(\ref{reg-02}) 
is not surprising since $\la\bar\psi\psi\ra_{\rm vac}$ is the sigma-term 
of the vacuum (per unit volume, up to the explicit factor of $m$).
Thus in the $\chi$QSM the picture emerges that $\sigmaPiN$ 
(measure of chiral symmetry breaking in the nucleon) 
is directly proportional to $\la\bar\psi\psi\ra_{\rm vac}$ 
(measure of spontaneous chiral symmetry breaking) and $m$ 
(measure of explicit chiral symmetry breaking).
The relation (\ref{propto-vac-cond}) is also known from the Skyrme model
\cite{Adkins:1983hy,Donoghue:1985bu} which is not surprizing, see the 
remarks in the previous paragraph.

The proportionality factor in Eq.~(\ref{propto-vac-cond}), the function 
$B(t)$ at $t=0$ as defined in (\ref{eval-05b}), has the dimension of volume. 
Following the temptation to define ``an effective volume of the nucleon'' as 
$V_{\rm eff}\equiv B(0)$, the relation between $\sigmaPiN$ and 
$\la\bar\psi\psi\ra_{\rm vac}$ can be written as
\be\label{int-01}
	\sigmaPiN = - m\,\la\bar\psi\psi\ra_{\rm vac}\;V_{\rm eff} \;,
	\;\;\; V_{\rm eff}\equiv B(0) .\ee
Numerically we find an effective volume which -- taking the nucleon to 
be a rigid sphere -- would correspond to an effective nucleon radius 
of $0.9\,{\rm fm}$ (for $m_\pi=140\,{\rm MeV}$).
This value should not be taken too seriously. But it yields the correct 
order of magnitude for the phenomenological size of the nucleon.
$V_{\rm eff}$ has a well defined chiral limit, see below.
The concept of such an effective volume is useful, e.g., in the context 
of the ``partial restoration of chiral symmetry in nuclear matter''.
The $V_{\rm eff} \equiv \sigmaPiN/(m_\pi^2f_\pi^2)$ in Eq.~(\ref{int-01})
is just the inverse of the chiral nucleon density $\rho_N^\chi$ introduced 
in Ref.~\cite{Cohen:1991nk}.

Thus, the value of $\sigmaPiN$ is large 
because $(-m\la\bar\psi\psi\ra_{\rm vac})$ is sizable {\sl and} 
because the nucleon is a large extended object. Corrections to 
this picture are suppressed by the instanton packing fraction.

\paragraph*{The strangeness content of the nucleon.}
Our result $\sigmaPiN=68\,{\rm MeV}$ implies a strangeness content in 
the nucleon of $y\sim 0.4$, cf.\  Section~\ref{Sec:sigma-general}, 
as it is also inferred from the more recent data analyses 
\cite{Kaufmann:dd,Olsson:1999jt,Pavan:2001wz,Olsson:pi}.
In view of such a large value for the strangeness content some 
authors even suggest to reconsider the ``standard interpretation'' 
relating $\sigmaPiN$ and $y$ \cite{Pavan:2001wz}.
(However, as argued in \cite{Ji:1994av} -- despite the large strangeness 
content -- the {\sl total} contribution of the strange degree of freedom 
to the nucleon mass is small.)

The calculation presented in this work definitely confirms -- within its
accuracy -- that $\sigmaPiN$ is large, and suggests an ``explanation'' why
(``because the nucleon is a large extended object'', see previous paragraph). 
Being a calculation in a SU(2) model, however, it unfortunately cannot
provide any insights into the puzzle why the strangeness content $y$ 
appears to be so large.
In this context it is worthwhile mentioning that in Ref.~\cite{Kim:1995hu},
where $\sigmaPiN$ was studied in the SU(3) version of the $\chi$QSM and
the matrix element $\la N|\bar\psi_s\psi_s|N\ra$ directly evaluated, 
the strangeness content was determined to be $y\sim 0.3$. 

The numerical result of Ref.~\cite{Kim:1995hu} does not provide any physical 
intuition why this value is so large. It would be interesting to see whether 
the present approach could be extended to the SU(3) sector and provide any 
insigths in this respect. Such a study, however, goes beyond the scope of 
this work.

\section{\boldmath Pion mass dependence of $\sigmaPiN$ and $\Mn$}
\label{Sec:chiral-limit}

The result in Eq.~(\ref{reg-06}) allows us to study the form factor 
$\sigma(t)$ as a function of $m_\pi$, i.e. $\sigma(t,m_\pi)$.
Of particular interest is hereby the behaviour in the chiral limit
which will be investigated first.
Maybe more interesting from a phenomenological point of view is the 
$m_\pi$-dependence of the nucleon mass $\Mn$ which can be deduced from
$\sigmaPiN(m_\pi)$ by means of the Feynman-Hellmann theorem.
The understanding of the correlation between $\Mn$ and $m_\pi$ is presently of 
importance for extrapolations of lattice data -- which correspond to pion
masses of typically $m_\pi\gtrsim 500\,{\rm MeV}$ -- to the physical value 
of the pion mass.

\paragraph*{Leading non-analytic contributions.}
Expanding the result in Eq.~(\ref{reg-06}) around the point $m_\pi=0$ 
we obtain for $\sigmaPiN(m_\pi)$,
$\Delta_\sigma(m_\pi)\equiv\sigma(2m_\pi^2,m_\pi)-\sigmaPiN(m_\pi)$ 
and the derivative $\sigma'(0,m_\pi)$ the following results
\ba
	\sigmaPiN(m_\pi) &=& 
	a_0\;m_\pi^2-\frac{27}{64\pi}\;\frac{g_A^2}{f_\pi^2}\;m_\pi^3+\dots\;,
	\label{Eq:ChiLim-02a} \\
	\Delta_\sigma(m_\pi) &=& 
   	\frac{9}{64\pi}\;\frac{g_A^2}{f_\pi^2}\; m_\pi^3 + \dots\;,
	\label{Eq:ChiLim-02b}  \\
	\sigma'(0,m_\pi) &=& 
	\frac{9}{256\pi}\;\frac{g_A^2}{f_\pi^2}\; m_\pi + \dots\;,
	\label{Eq:ChiLim-02c}\ea 
where the constant $a_0 = (3\sqrt{6\pi}g_A^{3/2})/(32f_\pi)$ and the dots
denote higher order terms in the limit $m_\pi\to0$ 
(see Appendix~\ref{App:sigma(t)}).
Since $m_\pi^2 \propto m$, odd powers of $m_\pi$ in 
Eqs.~(\ref{Eq:ChiLim-02a},~\ref{Eq:ChiLim-02b},~\ref{Eq:ChiLim-02c}) are
the respectively leading non-analytic contributions in the current 
quark mass $m$. 
Leading non-analytic contributions are of particular interest,
because they are model-independent. The leading non-analytic contributions 
in Eqs.~(\ref{Eq:ChiLim-02a},~\ref{Eq:ChiLim-02b},~~\ref{Eq:ChiLim-02b}) are 
exactly three times larger than those obtained in chiral perturbation theory. 
The discrepancy is explained by recalling that here we work in the 
large-$N_c$ limit and that the limits $N_c\to\infty$ and $m_\pi\to 0$
do not commute \cite{Gasser:1980sb,Dashen:1993jt}. In the large-$N_c$ 
counting $M_\Delta-\Mn={\cal O}(N_c^{-1})$ while $m_\pi={\cal O}(N_c^0)$
such that it is appropriate to consider first the large-$N_c$ limit, which is 
done here (though $M_\Delta-M_N>m_\pi$ in nature would suggest the opposite).
$\Mn(m_\pi)$ has a branching point at $(m_q\propto)\;m_\pi^2=(M_\Delta-\Mn)^2$ 
which in the strict large-$N_c$ limit contributes to the non-analytic behaviour
of $\Mn(m_\pi)$ at $m_\pi^2=0$.
As a consequence the $\Delta$ contributes as intermediate state in chiral 
loops to leading non-analytic contributions, and its contribution is twice 
(for isoscalar quantities) that of intermediate nucleon states in the strict 
large $N_c$ limit \cite{Cohen:1992uy}.

The result in Eq.~(\ref{Eq:ChiLim-02c}) means that the scalar isoscalar 
mean square radius diverges in the chiral limit
\be\label{r_S^2-in-chi-lim}
	\la r_S^2\ra = \biggl(\frac{27 g_A}{32\pi}\biggr)^{1/2}\;
	\frac{1}{m_\pi f_\pi} \cdot
	\biggl\{1 + {\cal O}(\mbox{regular terms})\biggr\}\;.
\ee
Another example of a square radius which diverges (however, as $\ln m_\pi$) 
in the chiral limit is the electric isovector charge mean square radius. 
(Also this feature is observed in the $\chi$QSM \cite{Christov:1995vm}).

Interestingly, the correct leading non-analytic contributions follow here
from the structure of the soliton, and not from chiral loops as in chiral
perturbation theory. 
In Ref.~\cite{Cohen:1992uy} it was shown that the leading non-analytic 
contribution to $\sigmaPiN$ in Eq.~(\ref{Eq:ChiLim-02a}) is a general 
result in a large class of chiral soliton models of the nucleon.
This result is reproduced here in the $\chi$QSM because the 
analytic profile (\ref{arctan-profile}) correctly describes the 
long distance behaviour of the chiral pion field \cite{Cohen:1992uy}.

%
%
\begin{figure*}[t!]\label{fig2:sigmaR}
\begin{tabular}{cc}
	\includegraphics[width=7.5cm,height=6.5cm]{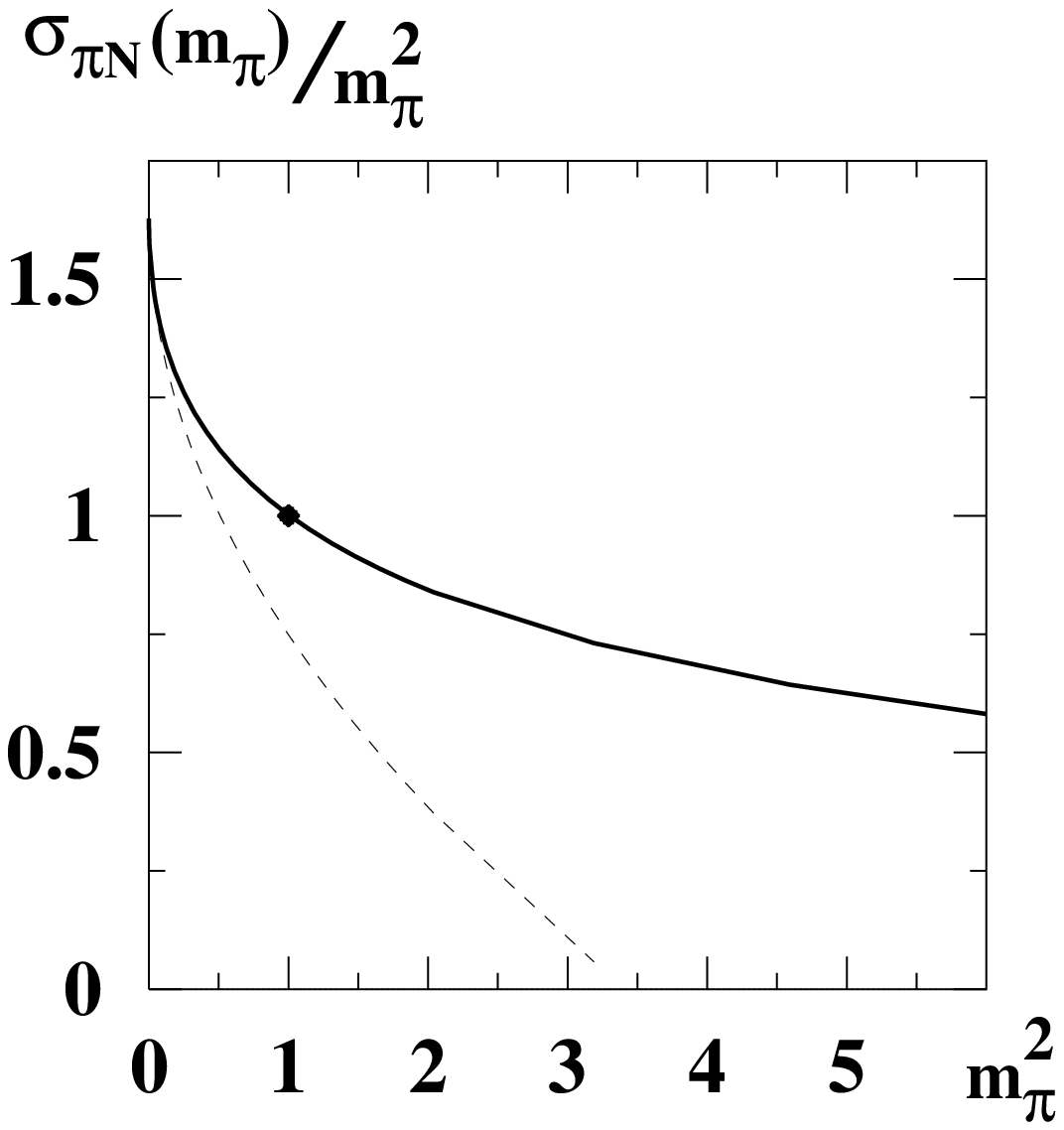} &
	\includegraphics[width=8cm,height=6.5cm]{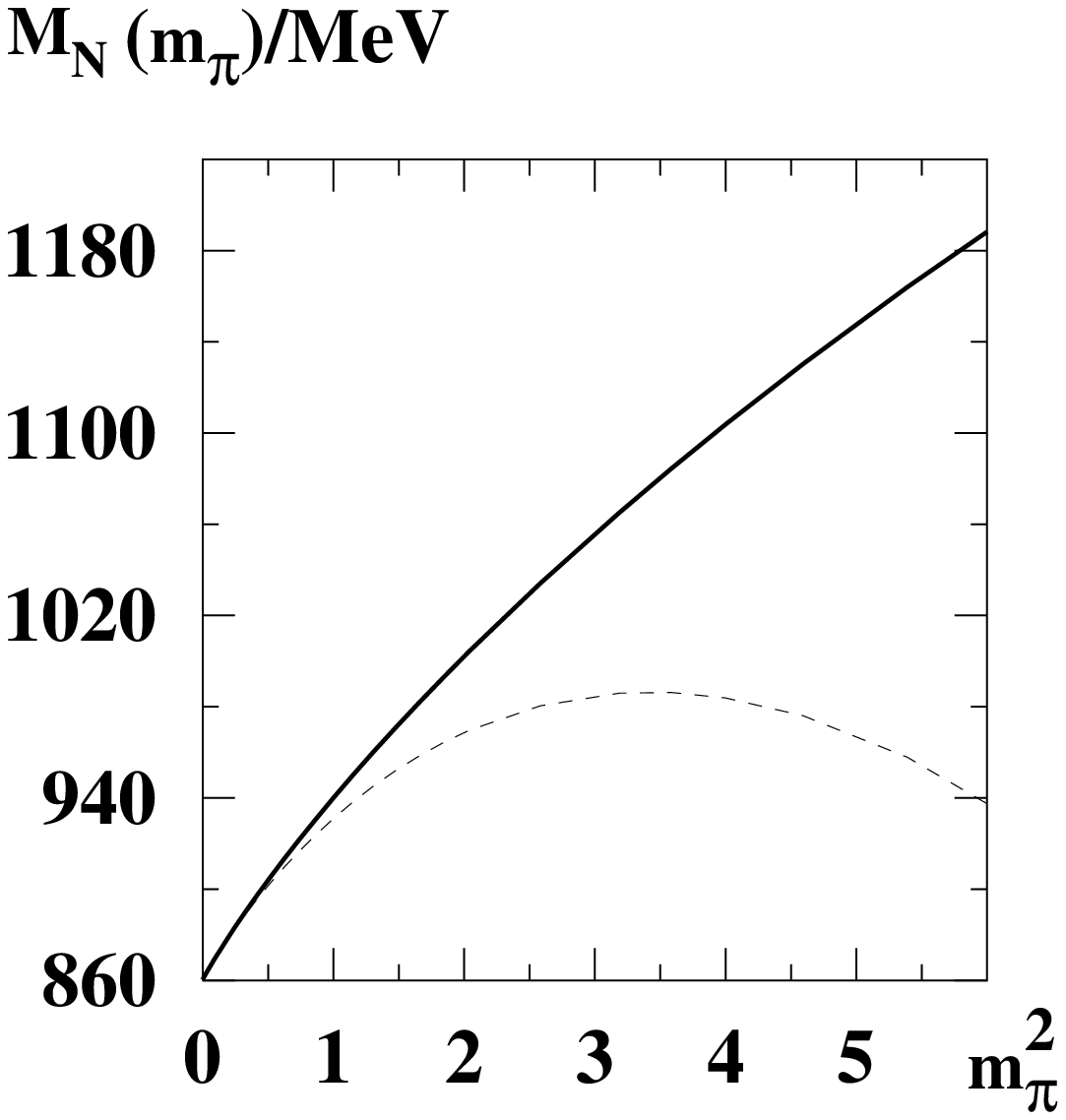}
	\cr
	{\bf a} & 
	{\bf b} 
\end{tabular}
	\caption{\footnotesize
	{\bf a.}
	The ratio $\sigmaPiN(m_\pi)/m_\pi^2$ as function of $m_\pi^2$ (with 
	$m_\pi$ in units of the physical pion mass and $\sigmaPiN(m_\pi)$ in
	in units of its physical (model) value $\sigmaPiN=68\,{\rm MeV}$, 
	i.e.\  the marked point corresponds to the physical situation).
	The solid line is the full model result, the dashed line is the 
	chiral expansion in the model up to the leading non-analytic 
	contribution, cf.\  Eq.~(\ref{Eq:ChiLim-02a}).
	{\bf b.} 
	$\Mn(m_\pi)$ as function of $m_\pi^2$ (in units of the physical 
	pion mass). Solid line is the full result, Eq.~(\ref{Eq:ChiLim-04}). 
	Dashed line is the (large-$N_c$) chiral expansion of $\Mn(m_\pi)$ 
	up to the leading non-analytic contribution, i.e.\  
	Eq.~(\ref{Eq:ChiLim-02a}) integrated by means of (\ref{Eq:ChiLim-03}).}
\end{figure*}
%
%

\paragraph*{Full dependence on $m_\pi$.}
Since $\sigmaPiN(m_\pi)$ vanishes in the chiral limit as $m_\pi^2$ 
we consider the ratio $\sigmaPiN(m_\pi)/m_\pi^2$. 
Figure~2a shows $\sigmaPiN(m_\pi)/m_\pi^2$ as function of $m_\pi^2$.
It is preferable to plot $m_\pi^2$ on the x-axis since the chiral limit 
corresponds to current quark mass $m\propto m_\pi^2 \to 0$. 
The solid line shows the full result and the dashed line shows the expansion 
of $\sigmaPiN(m_\pi)/m_\pi^2$ in the model up to the leading non-analytic 
contribution, Eq.~(\ref{Eq:ChiLim-02a}).
Clearly, only for rather small pion masses $m_\pi \ll 140\,{\rm MeV}$
the full result for $\sigmaPiN(m_\pi)/m_\pi^2$ is reasonably approximated 
by its chiral expansion up to the leading non-analytic contribution in 
Eq.~(\ref{Eq:ChiLim-02a}). 

Figure~2 simultaneously shows the ``effective volume'' of 
the nucleon $V_{\rm eff}(m_\pi)$ as defined in (\ref{int-01}) in units 
of its ``physical value'' (corresponding to about $3\,{\rm fm}^3$).
In the chiral limit $V_{\rm eff}(m_\pi)$ grows by more than $50\%$ 
compared to its physical value. But it remains finite which means 
that this quantity is indeed a useful measure of the ``nucleon size'' 
in the chiral limit -- in contrast to, e.g., the scalar mean square 
radius $\la r_S^2\ra$, cf.\  Eq.~(\ref{r_S^2-in-chi-lim}).

\paragraph*{Nucleon mass as function of $m_\pi$.}
The Feynman-Hellmann theorem (\ref{way2}), reformulated 
by means of the Gell-Mann--Oakes--Renner relation in Eq.~(\ref{GMOR}) as
\be\label{Eq:ChiLim-03}
	\sigmaPiN(m_\pi) = m_\pi^2\frac{\partial\Mn(m_\pi)}{\partial m_\pi^2}
	\;,\ee
provides a mean to study the nucleon mass as function of $m_\pi$
\be\label{Eq:ChiLim-04}
	\Mn(m_\pi) = \Mn(0) + \int_0^{m_\pi^2} \!\!\di\mu^2\;
	\frac{\sigmaPiN(\mu)}{\mu^2} \;, \ee
where $\Mn(0)$ is an integration constant to be identified with the 
value of the nucleon mass in the chiral limit.

For the amount the nucleon mass is shifted in the chiral limit with respect
to its physical value we obtain from (\ref{Eq:ChiLim-04}) the result 
\be\label{Eq:ChiLim-05}
	\Mn(140\,{\rm MeV}) - \Mn(0)
	= \int_0^{m_\pi^2}\!\! \di\mu^2\;\frac{\sigmaPiN(\mu)}{\mu^2} 
	  \biggl|_{m_\pi = 140\,{\rm MeV}} 
	= 80\,{\rm MeV} \; \ee
modulo corrections which are parametrically small in the instanton packing 
fraction. Within this accuracy the mass of the nucleon in the chiral limit 
is $\Mn(0)=860\,{\rm MeV}$ (taking the physical mass as $940\,{\rm MeV}$),
which is in the range of the values considered in chiral perturbation theory.
Since we work here in the large-$N_c$ limit one obtains the same result for 
the mass shift of the $\Delta$ in the (large-$N_c$) chiral limit.

Fixing the integration constant $\Mn(0)$ in (\ref{Eq:ChiLim-04}) to 
$860\,{\rm MeV}$ we obtain for $\Mn(m_\pi)$ the result shown in Fig.~2b, 
where the solid line shows the full result from Eq.~(\ref{Eq:ChiLim-04}) 
and the dashed line shows the chiral expansion of $\Mn(m_\pi)$  
up to the leading non-analytic contribution,
Eqs.~(\ref{Eq:ChiLim-02a},~\ref{Eq:ChiLim-04}). 
Similarly to the case of $\sigmaPiN(m_\pi^2)$ the chiral expansion up to 
the leading non-analytic contribution reasonably approximates the full 
result for $\Mn(m_\pi)$ only for $m_\pi$ below the physical pion mass.

%
%
\begin{figure*}[t!]
\begin{tabular}{cc}
	\includegraphics[width=7.5cm,height=6.5cm]{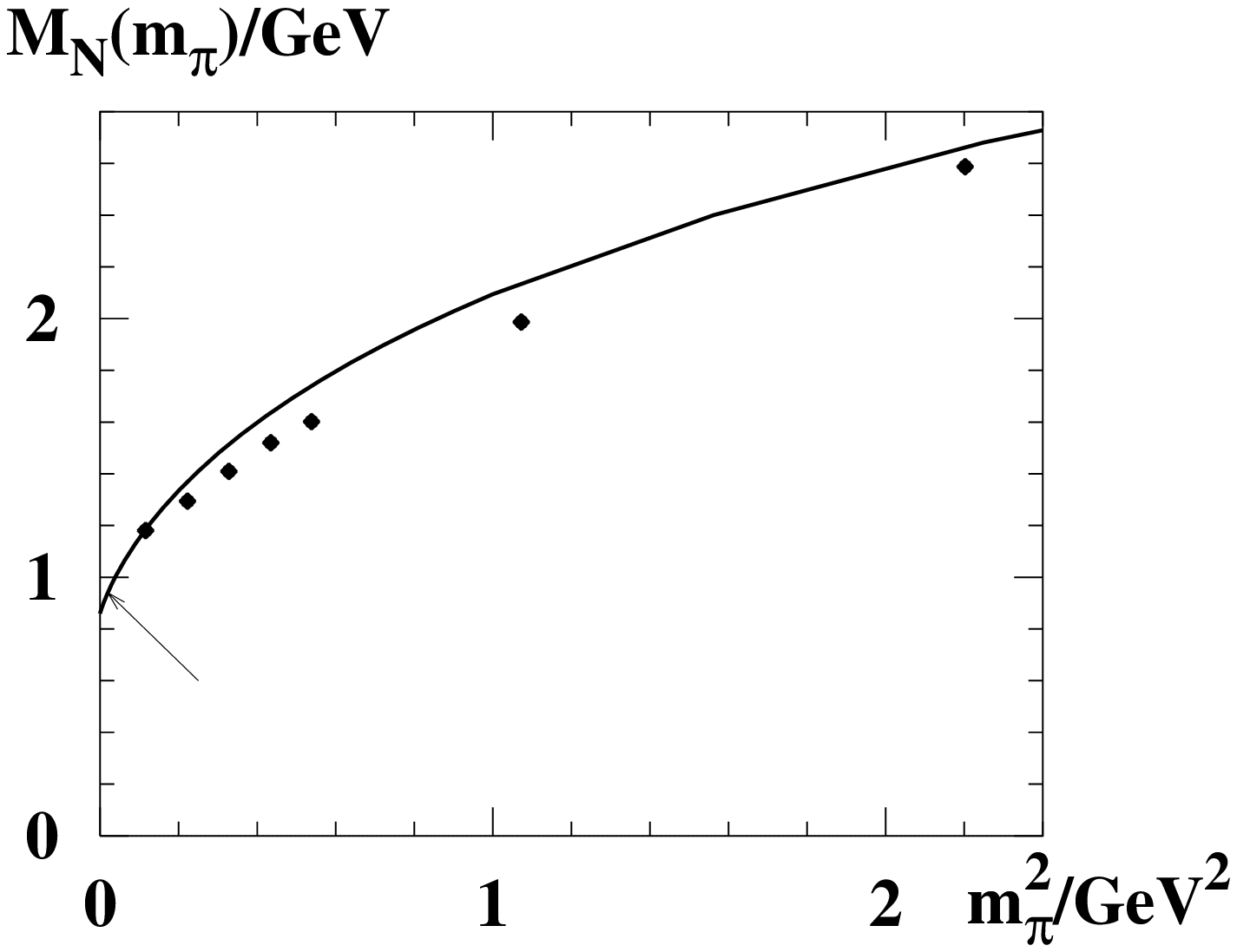} &
	\includegraphics[width=7.5cm,height=6.5cm]{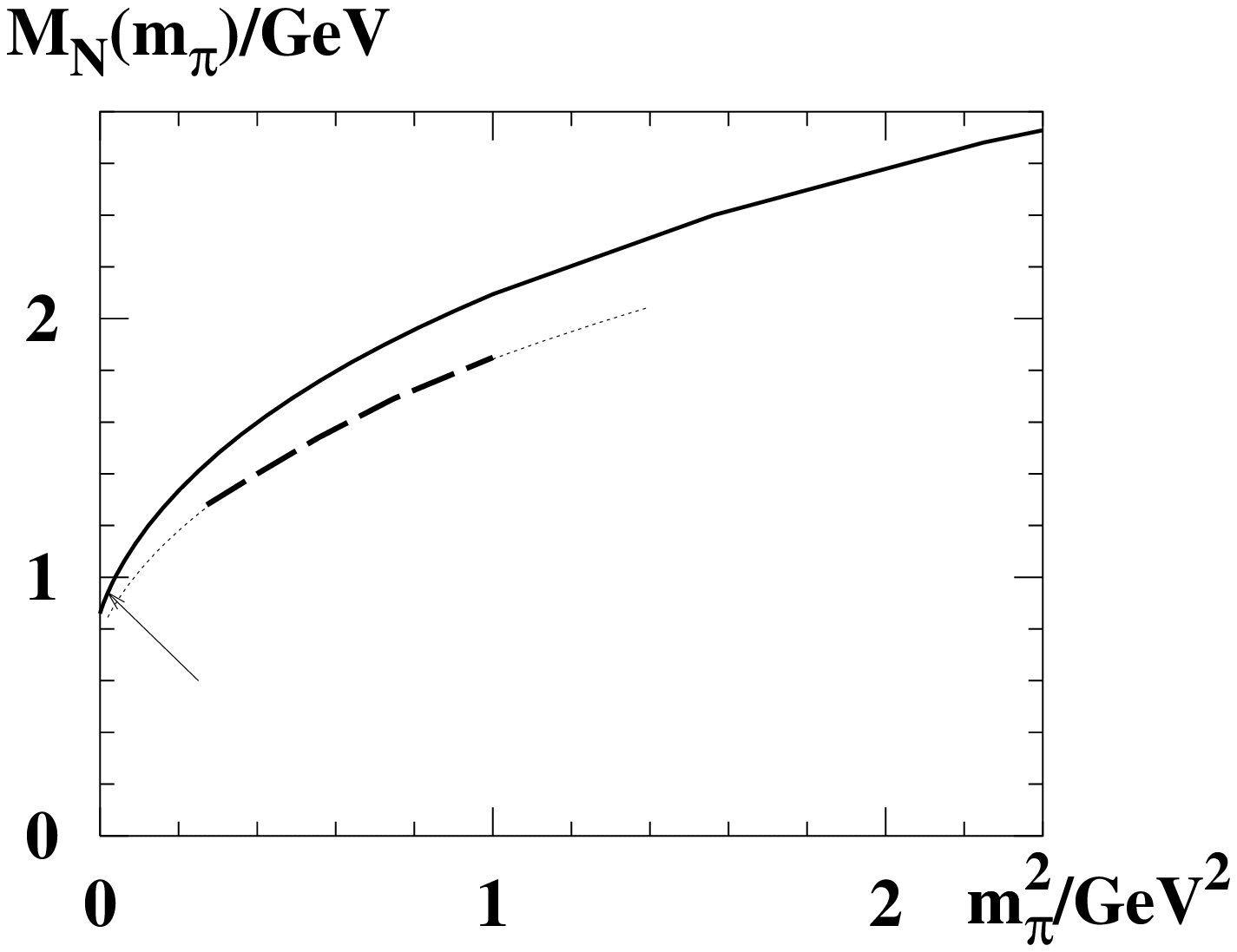}
	\cr
	{\bf a} & 
	{\bf b} 
\end{tabular}
	\caption{\footnotesize
	The nucleon mass $\Mn$ as a function of $m_\pi^2$ according to
	Eq.~(\ref{Eq:ChiLim-04}) with $\Mn(0)=860\,{\rm MeV}$ in comparison 
	to ({\bf a}) lattice QCD results from \cite{Bernard:2001av}, and
	({\bf b}) from \cite{AliKhan:2001tx}. (The thick dashed line is
	actually the fit of Ref.~\cite{Leinweber:2000sa} to the lattice 
	results \cite{AliKhan:2001tx}.)
	The meaning of the thin dotted line is explained in the text.
	In both figures the arrow denotes the physical point.}
\end{figure*}
%
%

Several comments are in order. 
First, in the $\chi$QSM the nucleon mass is given by the minimum of the 
soliton energy (\ref{soliton-energy}) with respect to variations of the 
chiral field $U$. 
The accurate (and numerically involved) procedure to study the {\sl exact} 
dependence of $\Mn$ on the pion mass $m_\pi$ in the $\chi$QSM would consist 
in considering the soliton energy (\ref{soliton-energy}) as a function of the 
current quark mass $m$ in Eqs.~(\ref{Hamiltonian},~\ref{free-Hamiltonian}) and 
deducing the respective value of the pion mass from the 
Gell-Mann--Oakes--Renner relation (\ref{GMOR}), or the Yukawa-like decay of 
the self-consistent soliton profile $\propto \exp(-m_\pi r)/r$ at large $r$. 
Here, however, we can be confident to describe correctly the {\sl variation} 
of $\Mn$ with $m_\pi$ to within the accuracy of $\sigmaPiN$ in 
Eq.~(\ref{reg-06}).

Next, since Eq.~(\ref{Eq:ChiLim-04}) can be used to describe merely the 
variation of $\Mn$ with $m_\pi$ it is clear that we are free to choose the
integration constant $\Mn(0)=860\,{\rm MeV}$ and that there would be no point 
in taking the precise model value for $\Mn(0)$ which, by the way, would have 
the drawback of introducing a regularization scheme dependence into the result.
It should be noted that in the $\chi$QSM the baryon masses tend to be 
overestimated by about $20\%$ due to spurious contributions of the center 
of mass motion of the soliton \cite{Pobylitsa:1992bk}.

Finally, one has to consider that the description of $\Mn(m_\pi^2)$ by means 
of Eq.~(\ref{Eq:ChiLim-04}) can be considered as well justified for 
$m_\pi^2\rho_{\rm av}^2 \ll 1$. 
This relation is analog to Eq.~(\ref{packing-fraction}) and means that only 
light pions -- light with respect to the scale $\rho_{\rm av}^{-1}$ -- make 
sense as effective degrees of freedom in the low energy theory 
(\ref{eff-theory}).

Though, of course, it is not their actual goal, lattice QCD simulations
allow to measure $\Mn(m_\pi)$. The result for $\Mn(m_\pi)$ from 
Eq.~(\ref{Eq:ChiLim-04}) with $\Mn(0)=860\,{\rm MeV}$ is compared to results
from lattice QCD simulations from Refs.~\cite{Bernard:2001av,AliKhan:2001tx}
in Figs.~3a and 3b. (More precisely, what is plotted in Fig.~3b as a thick
dashed line is the parameterization of the lattice data \cite{AliKhan:2001tx}
reported in Ref.~\cite{Leinweber:2003dg}.)

Keeping in mind the above-mentioned reservations we observe in Fig.~3 a good 
agreement with the results for the nucleon mass from lattice QCD simulations 
reported in Refs.~\cite{Bernard:2001av,AliKhan:2001tx}.
A similarly good agreement is observed with the results reported by other 
lattice groups \cite{Zanotti:2001yb,Aoki:1999ff}.

\paragraph*{Chiral extrapolation of lattice-QCD results.}
Eventually one is interested in the chiral 
extrapolation of lattice data from the currently available region 
$500\,{\rm MeV} \lesssim m_\pi \lesssim \mbox{several}\,{\rm GeV}$ 
to the physical value of the pion mass.
The lattice data on $\Mn(m_\pi)$ are usually fitted to ans\"atze of the kind 
$\Mn(m_\pi) = a + b m_\pi^2$ or $\Mn(m_\pi) = a + b m_\pi^2 + c m_\pi^3$
inspired by the chiral expansion of the nucleon mass, often observing that 
both ans\"atze are equally acceptable \cite{Aoki:1999ff}.
However, the approximation of $\Mn(m_\pi)$ by its chiral expansion up to 
the leading non-analytic term ${\cal O}(m_\pi^3)$ makes sense only at small 
values of $m_\pi$ below the physical pion mass -- as was emphasized in 
\cite{Leinweber:2000sa} (and can also be seen here in Fig.~2b).

An interesting extrapolation method  was introduced in \cite{Leinweber:1999ig}
where it was suggested to regularize chiral pion loops (which simulate the 
pion cloud effect) by introducing appropriate regulators (i.e. form factors 
which simulate the extended structure of the nucleon). 
This approach not only incorporates the correct chiral behaviour, but 
also reproduces results of the heavy quark effective theory in the limit
$m\propto m_\pi^2\to\infty$. This gives a certain legitimation that also 
the intermediate $m_\pi$-region -- as explored in lattice QCD calculations 
-- is reasonably described in this approach.
The model dependence of this approach was studied in 
\cite{Leinweber:2000sa,Leinweber:2003dg} by using different regulators.
The thick dashed curve in Fig.~3b is the fit of Ref.~\cite{Leinweber:2003dg}
to the lattice data \cite{AliKhan:2001tx} in the range 
$500\,{\rm MeV}\lesssim m_\pi \lesssim 1\,{\rm GeV}$.
(The thickness of the curve is comparable to the statistical error 
of the very accurate lattice data of Ref.~\cite{AliKhan:2001tx}.)
The extrapolation from this region to the physical pion mass yields --
within the statistical accuracy of the lattice data and depending on the
regulator -- values for the nucleon mass which cover the region
$(782^{+122}_{-122} \dots 948^{+\;80}_{-216})\,{\rm MeV}$.
This demonstrates the sensitivity to details of the extrapolation 
of even very accurate lattice data. 

The approach of 
Refs.~\cite{Leinweber:2000sa,Leinweber:2003dg,Leinweber:1999ig} basically 
corresponds to chiral perturbation theory in different regularization schemes.
From the point of view of field theory, however, it is unsatisfactory 
to observe a strong scheme dependence \cite{Bernard:2003rp}.
Considering the complexity of the problem it is important to have further 
(model independent) constraints. Recently it was reported that chiral 
perturbation theory is able to describe reliably $\Mn(m_\pi)$ up to 
$m_\pi < 600\,{\rm MeV}$ using refined regularization techniques
\cite{Bernard:2003rp}. Presently most of the lattice data is beyond 
that limit, however, a first matching of chiral perturbation theory 
and lattice results seems to be possible \cite{Bernard:2003rp}.

The results reported in this work also could provide useful
-- though certainly not model-independent -- insights into this issue.
Indeed, by reasonably fixing the integration constant $\Mn(0)$,
an agreement with lattice data to within the accuracy of the approach 
is obtained, cf.\  Figs.~3a and 3b. 
This observation suggests that it would be worthwhile attempting 
to fit lattice data with an ansatz like, e.g.,
\be\label{ansatz}
	\Mn(m_\pi) = \Mn(0)^{\rm fit} + \int\limits_0^{\;\;m_\pi^2}\!
	\frac{\di\mu^2}{\mu^2}\;\sigmaPiN(\mu,R_{\rm sol}^{\rm fit})\;.
\ee
The ansatz (\ref{ansatz}) corresponds to Eq.~(\ref{Eq:ChiLim-04}) where
$\Mn(0)$ and the soliton size $R_{\rm sol}$ (cf.\ Eq.~(\ref{arctan-profile}))
are allowed to be free parameters, $\Mn(0)^{\rm fit}$ and 
$R_{\rm sol}^{\rm fit}$, to be fitted to lattice data.
The ansatz (\ref{ansatz}) is of more general character:
It is not only inspired by the $\chi$QSM ({\sl and} the Skyrme model, cf.\  
previous Section~\ref{Sec:results+discussion}) but actually is based on the 
large $N_c$ description of the nucleon as a chiral soliton of the pion field 
\cite{Witten:tx,Adkins:1983ya}.

The ansatz (\ref{ansatz}) means that the pion cloud effect -- which is 
responsible for the growth of the nucleon mass with increasing $m_\pi$ -- 
is modelled by the structure of the soliton. 
This is conceptually an independent ansatz to consider pion mass effects than 
the method of Refs.~\cite{Leinweber:2000sa,Leinweber:2003dg,Leinweber:1999ig},
and it has moreover the advantage of being regularization scheme independent.

In order to illustrate that the ansatz (\ref{ansatz}) is reasonable 
note that $\Mn(0)^{\rm fit} = 780\,{\rm MeV}$ and
$R_{\rm sol}^{\rm fit} = 0.93 R_{\rm sol}$ fit the lattice data 
\cite{AliKhan:2001tx} within their statistical accuracy.
The ansatz (\ref{ansatz}) with these parameters -- shown in Fig.~3b as the 
thin dotted line in comparison to the fit obtained in \cite{Leinweber:2003dg}
-- yields for the physical nucleon mass about $850\,{\rm MeV}$ which 
is in the range of the values reported in Ref.~\cite{Leinweber:2003dg}. 
A careful study whether this is the {\sl best} fit and an estimate of its 
statistical {\sl and} systematic errors go both beyond the scope of this 
work. The main systematic error in the ansatz (\ref{ansatz}) is due to 
the treatment of the $\Delta$ as a mass generated state to the nucleon 
in the large $N_c$ limit and could be estimated following 
Ref.~\cite{Cohen:1992uy}.

\section{Summary and Conclusions}
\label{Sec:conclusions}

The sigma-term form factor $\sigma(t)$ of the nucleon was studied 
in the limit of a large number of colours in the framework of the 
chiral quark-soliton model ($\chi$QSM). 
The consistency of this field theoretical approach is illustrated by the fact 
that the pion-nucleon sigma-term $\sigmaPiN$ can consistently be computed in
the $\chi$QSM in three different ways.
Apart from continuing analytically the form-factor to $t=0$
one also can make use of the Feynman-Hellmann theorem 
\cite{Feynman-Hellmann-theorem} and the sum rule for the twist-3 
distribution function $e^a(x)$ \cite{Jaffe:1991kp}.

The model expression for $\sigma(t)$ was evaluated in an approximation 
justified by arguments from the instanton model of the QCD vacuum from
which the $\chi$QSM was derived. The approximation is therefore 
theoretically well controlled and justified. A virtue of the approximation 
is that it allows to regularize $\sigma(t)$, which is quadratically 
UV-divergent in the $\chi$QSM, in a regularization scheme independent way. 
The theoretical accuracy of the results is governed by the smallness of the
parameter characterizing the diluteness of the instanton medium and is
practically of ${\cal O}(30\%)$. Results from previous exact calculations 
in the $\chi$QSM are reproduced to within this accuracy.
There are no adjustable parameters in the approach.

For the form factor at the Cheng-Dashen point the value 
$\sigma(2m_\pi^2)=83\,{\rm MeV}$ is found -- in good agreement
with more recent analyses of pion-nucleon scattering data
\cite{Kaufmann:dd,Olsson:1999jt,Pavan:2001wz,Olsson:pi}.
In the region $-2m_\pi^2 < t < 4m_\pi^2$ the result for the form factor 
agrees to within few percent with the shape for $\sigma(t)$ obtained in 
\cite{Gasser:1990ce} on the basis of a dispersion relation analysis.
In particular $\sigma(2m_\pi^2)-\sigma(0)=14.7\,{\rm MeV}$ is obtained 
which is close to the value $15.2\,{\rm MeV}$ of Ref.~\cite{Gasser:1990ce}.
Finally, for the pion-nucleon sigma-term the value $\sigmaPiN=68\,{\rm MeV}$ 
is found.

An advantage of the present approximate result for $\sigma(t)$ is its simple
structure which allows, e.g., to study the form factor in the chiral limit 
and to derive the leading non-analytic (in the current quark mass) 
contributions in the model. It was shown that the expressions from the
$\chi$QSM contain the correct leading non-analytic contributions in the 
large-$N_c$ limit \cite{Cohen:1992uy}.

The results presented in this work fully confirm the observation made 
in other chiral models that the dominant contribution to $\sigmaPiN$ is due
to the pion cloud. Indeed, here the extreme situation emerges that $\sigmaPiN$
is solely due to the pion cloud -- modulo small corrections 
suppressed by the instanton packing fraction.
Moreover it is found that $\sigmaPiN$ is proportional to the quark vacuum 
condensate, i.e.\  the order parameter characterizing spontanous chiral 
symmetry breaking, and to the current quark mass, i.e.\  the parameter 
characterizing explicit chiral symmetry breaking -- a relation previously
known only from the Skyrme model. 
The proportionality factor can be interpreted as an effective volume of
the nucleon, which yields the physically appealing explanation that 
$\sigmaPiN$ is so sizeable because the nucleon is a large extended object.

The large value of $\sigmaPiN$ implies according to the standard interpretation
a surprizingly sizeable strangeness content of the nucleon of about $y\equiv$ 
$2\la N|\bar\psi_s\psi_s|N\ra/\la N|\bar\psi_u\psi_u+\bar\psi_d\psi_d|N\ra$
$\sim 0.4$. The SU(2) calculation presented here unfortunately cannot shed 
any light on this puzzle. Calculations in the SU(3) version of the $\chi$QSM, 
however, naturally accomodate a large strangeness content \cite{Kim:1995hu}. 
Worthwhile mentioning in this context is the conclusion \cite{Ji:1994av} that
despite the large value of $y$ the total contribution of strange quarks to 
the nucleon mass is small.

The dependence of the nucleon mass $\Mn$ on the pion mass was obtained 
from $\sigmaPiN(m_\pi)$ by exploring the Feynman-Hellmann theorem. 
In the chiral limit the nucleon mass was found to be reduced by 
$80\,{\rm MeV}$ with respect to its physical value. This means that the 
mass of the nucleon in the chiral limit is about $860\,{\rm MeV}$ which is 
in the range of the values considered in chiral perturbation theory.
To within the accuracy of the approach the functional dependence of the 
nucleon mass on the pion mass agrees with lattice QCD results up to the 
largest available lattice values of $m_\pi$. This observation can be used 
to inspire ans\"atze for the chiral extrapolation of lattice QCD results.

Despite the considerable progress of the rigorous approaches to QCD 
 -- lattice QCD and the effective field theory method of chiral
perturbation theory -- the description of the pion-nucleon sigma-term 
still is a demanding issue. The direct calculation of $\sigmaPiN$ on 
the lattice meets the problem that the corresponding operator is not 
renormalization scale invariant \cite{Dong:1995ec}. 
This complication is avoided by the method of determining 
$\sigmaPiN$ from the pion mass dependence of the nucleon mass 
\cite{Bernard:2001av,AliKhan:2001tx,Zanotti:2001yb,Aoki:1999ff}.
In either case, however, the conceptual problem appears of how to extrapolate 
lattice data from the presently available $m_\pi\gtrsim 500\,{\rm MeV}$ down 
to the physical value of the pion mass. 
A model-independent guideline can, in principle, be delivered by the 
chiral perturbation theory where, however, the physical value of $\sigmaPiN$ 
cannot directly be computed since it serves to absorbe counter terms and is
renormalized anew in each order of the chiral expansion.

Recently a first promising matching of chiral perturbation theory to 
the lowest presently available pion masses in lattice QCD was reported 
\cite{Bernard:2003rp}. However, it will take some time to demonstrate 
and establish the convergence of the chiral expansion up to values of 
$m_\pi\sim 500\,{\rm MeV}$ and higher.
Until then less rigorous but phenomenologically and theoretically well 
motivated approaches may have a chance to serve as helpful guidelines for 
the chiral extrapolation of lattice data. One such approach, suggested in 
Ref.~\cite{Leinweber:1999ig}, consists in exploring physically motivated
regularization technics to simulate pion cloud effects. 
In order to investigate the model dependence it would be desirable to consider 
also other conceptually different approaches. 

The observations made in this work inspire an alternative, regularization 
scheme independent, extrapolation ansatz in which the pion cloud is modelled 
on the basis of the soliton picture of the nucleon. In view of the success of
this picture in describing baryon properties such ans\"atze appear to be 
promising. To provide a theoretically well controlled extrapolation ansatz 
it will, however, be necessary to introduce systematically $1/N_c$-corrections
which is subject to current investigations.

It would be interesting to study also other observables of the nucleon in 
this way, using arguments from the instanton vacuum  model -- which not 
only largely simplifies the calculations but is crucial, e.g., to allow 
the derivation of leading non-analytic contributions in the $\chi$QSM.
However, it is questionable whether hereby one could obtain also a
{\sl phenomenologically} satisfactory description of observables 
other than quadratically UV-divergent, and the only observable of 
such kind as far as the author is aware of is the scalar form factor.

\begin{acknowledgments}
The author thanks S.~Boffi, K.~Goeke and M.~V.~Polyakov for fruitful 
discussions. This work has partly been performed under the contract  
HPRN-CT-2000-00130 of the European Commission.
\end{acknowledgments}

\appendix
   \setcounter{equation}{0}
   \def\theequation{A.\arabic{equation}}

\section{Gradient expansion}
\label{App:gradient}

By using $1/(\omega+iH)=(\omega-iH)/(\omega^2+H^2)$ and 
$\hat{H}^2=\hat{\bf p}^2+M^2+iM\gamma^k(\nabla^k U^{\gamma_5})(\hat{\bf x})$
in Eq.~(\ref{eval-03}) we arrive at the series (\ref{eval-04}) with the
$\sigma(t)_{\rm cont}^{(k)}$ given by
\ba
	\sigma(t)_{\rm cont}^{(k)} &=& 
	m\;N_c \int_C \frac{\di\omega}{2\pi i} \;{\rm Sp} \Biggl[
	\,j_0(\sqrt{-t}\,|\hat{\bf x}|)\, \gamma^0 (\omega-i\hat{H})\;
	\frac{1}{\omega^2+\hat{\bf p}^2+M^2}\nonumber\\
	&& \mbox{\hspace{3cm}}\times 
	\biggl(-iM\gamma^k(\nabla^k U^{\gamma_5})(\hat{\bf x})\,
	\frac{1}{\omega^2+\hat{\bf p}^2+M^2}\biggr)^{\! k}
	- (\hat{H}\to\hat{H}_0)\;\Biggr]_{\rm reg}  \;. 
	\label{App:grad-01}\ea

In order to evaluate the zeroth order contribution 
$\sigma(t)_{\rm cont}^{(0)}$ we saturate the functional trace by the complete
set of eigenfunctions of the free momentum operator, i.e.\ 
${\rm Sp}[\dots]\equiv$ 
$\int\frac{\di^3{\bf p}}{(2\pi)^3}\la{\bf p}|{\rm tr}\dots|{\bf p}\ra$ 
where ${\rm tr}$ denotes the trace over flavour- and Dirac-indices.
After taking the Dirac-trace we obtain
\be\label{eval-05}
	\sigma(t)_{\rm cont}^{(0)} 
	=  -\;m\;N_c \; 4M \int\di^3{\bf x} \;j_0(\sqrt{-t}\,|\hat{\bf x}|)\;
	{\rm tr}_{\rm F}\biggl(\frac{U+U^\dag\!}{2} -1\biggr)
	\int_C\frac{\di\omega}{2\pi}\int\frac{\di^3{\bf p}}{(2\pi)^3}
	\,\frac{1}{\omega^2+{\bf p}^2+M^2}\biggl|_{\rm reg}\; \ee
The ``$-1$'' under the flavour-trace (${\rm tr}_{\rm F}$) originates from the 
vacuum subtraction term.  
The $\omega$-integration over the contour $C$ in (\ref{eval-05a}) can 
replaced by an integration over the real $\omega$-axis, and we substitute 
$\omega\to p^0_E$ to use the convenient Euclidean space notation.
Using ${\rm tr}_{\rm F}U ={\rm tr}_{\rm F}U^\dag$ we obtain the result 
in Eq.~(\ref{eval-04a}). 

Similarly we obtain for the $k=1$ term in (\ref{App:grad-01})
\be\label{eval-06}
	\sigma(t)_{\rm cont}^{(1)} 
	=\int_C \frac{\di\omega}{2\pi} 
	\int\frac{\di^3{\bf p}}{(2\pi)^3}
	\frac{i\;mN_cM\,8g_{ik}p^i}{(\omega^2+{\bf p}^2+M^2)^2}
	\int\di^3{\bf x} \;j_0(\sqrt{-t}\,|{\bf x}|)\,
	\nabla^k\cos P(|{\bf x}|) \biggr|_{\rm reg} 	
	= 0\;,\ee
which is zero due to rotational invariance.

The $k=2$ term in (\ref{App:grad-01}) yields after a 
somehow lengthy calculation
\be\label{eval-07}
	\sigma(t)_{\rm cont}^{(2)} = - \;\frac{mN_cM}{16\pi^2}
	\int\di^3{\bf x} \;j_0(\sqrt{-t}\,|{\bf x}|)\; \trF\, 
	U^\dag({\bf x})\,(\nabla^kU)({\bf x})\,(\nabla^kU^\dag)({\bf x})
	\;.\ee
For our purposes it is sufficient to observe that $\sigma(t)_{\rm cont}^{(2)}$
is finite, 
e.g., $(\sigmaPiN)_{\rm cont}^{(2)}/m \le -(3\sqrt{2}/16) N_cMR_{\rm sol}$.
(The equality holds for $m_\pi=0$ in the profile (\ref{arctan-profile}).) 
Higher orders $k\ge 3$ are also finite, i.e.\ 
\be\label{eval-08}
	\sigma(t)_{\rm cont}^{(k)} = \mbox{UV-finite} \;\;,\;\; 
	k \ge 2\;.\ee

   \setcounter{equation}{0}
   \def\theequation{B.\arabic{equation}}

\section{\boldmath $\sigma(t)$ and the chiral limit}
\label{App:sigma(t)}

Taking the trace over flavour-indices in the expression for $B(t)$ in 
(\ref{eval-05b}) and inserting the result into the expression (\ref{reg-06}) 
for $\sigma(t)$ we obtain for the soliton profile in (\ref{arctan-profile})
\ba
&&	\sigma(t) = m_\pi^2f_\pi^2 \; 4\pi \int\limits_0^\infty\di r\;r^2\;
	j_0(r\sqrt{-t})\,\biggl(1-\cos P(r)\biggr)  \;, \nonumber\\
&&  	1-\cos P(r) = \frac{2R_{\rm sol}^4(1+m_\pi r)^2\,\exp(-2m_\pi r)}
	{r^4+R_{\rm sol}^4(1+m_\pi r)^2\,\exp(-2m_\pi r)} \;.
	\label{discuss-01} \ea
The form-factor can be continued analytically to the phenomenologically 
interesting region $t\ge 0$ as
\be\label{discuss-02}
	j_0(r\sqrt{-t}) =\cases{
  \displaystyle\frac{\sin(r\sqrt{-t})}{r\sqrt{-t}}	& $t < 0$,\cr
  \phantom{\biggl|} 1 					& $t = 0$,\cr
  \displaystyle\frac{{\rm sinh}\,(r\sqrt{t})}{r\sqrt{t}}& $t > 0$.}
	\ee
The form factor (\ref{discuss-01}) is undefined for $t\ge 4m_\pi^2$.
For positive values of $t$ the integral over $r$ in (\ref{discuss-01}) 
converges only for $t< 4m_\pi^2$ because only then at large $r$ the decay
of $r^2[1-\cos P(r)]\propto\exp(-2m_\pi r)$ can 
compensate the rise of $j_0(r\sqrt{-t})\propto\exp(r\sqrt{t})/r$.

\paragraph*{Chiral limit.}
\label{App:sigmaPiN-in-chiral-lim}

The pion-nucleon sigma-term $\sigmaPiN(m_\pi)$ as function of $m_\pi$ 
is given by
\ba\label{int-02}
	\sigmaPiN(m_\pi) = 
	m_\pi^2 f_\pi^2 \; 2R_{\rm sol}^3 \;4\pi \;I(m_\pi R_{\rm sol})\;,
	\;\;\;
	I(a) \equiv \int\limits_0^\infty
	\frac{\di x\;x^2\,(1+ax)^2\,\exp(-2ax)}{x^4+(1+ax)^2\,\exp(-2ax)}
	\;.\ea
The zeroth order in the Taylor of the function $I(a)$ around $a=0$ is
\be\label{int-03}
	I(0) = \int\limits_0^\infty \frac{\di x\;x^2}{x^4+1}
	     = \frac{\pi}{2\sqrt{2}} \;. \ee
The linear term in the Taylor expansion of $I(a)$ requires careful treatment.
Consider $a\neq 0$ and introduce an upper limit $D\gg R_{\rm sol}$ in the
space integral in Eq.~(\ref{discuss-01}).
This simulates the ``realistic situation'' of computing $\sigmaPiN$ in a 
finite volume (as it happens in many model calculations and in lattice QCD).
Then
\be\label{int-04}
	\frac{\partial I(a)}{\partial a} = -2a \int\limits_0^{D/R_{\rm sol}}
	\frac{\di x\;x^8(1+ax)\exp(2ax)}{(x^4\exp(2ax)+(1+ax)^2)^2}
	= -2\int\limits_0^{Dm_\pi}
	\frac{\di z\;z^8(1+z)\exp(2z)}{(z^4\exp(2z)+a^4(1+z)^2)^2}
\ee
where we substituted $z=ax$ in the second step and used $a=m_\pi R_{\rm sol}$.
Taking the continuum limit $D\to\infty$ first, and only then $m_\pi\to 0$ 
(apperantly these limits do not commute) we arrive at
\be
	\frac{\partial I(a)}{\partial a} \;\stackrel{a\to 0}{\longrightarrow}
	\; -2\int\limits_0^\infty\di z\;(1+z)\exp(-2z) = - \frac{3}{2}\;.
\ee
Thus, we obtain
\be\label{int-05}
	I(a)=\frac{\pi}{2\sqrt{2}}-\frac{3}{2}\;a +
	{\cal O}(\mbox{higher orders in $a$})\;.
\ee
Inserting the result (\ref{int-05}) into Eq.~(\ref{int-02}) and 
making use of the relation (\ref{rel-gA-Rsol}) in order to eliminate  
$R_{\rm sol}$ in favour of the pion decay constant $f_\pi$ and nucleon axial 
coupling constant $g_A$ yields 
the result in Eq.~(\ref{Eq:ChiLim-02a}).
The results in Eqs.~(\ref{Eq:ChiLim-02b},~\ref{Eq:ChiLim-02c}) follow
in an analogous way.
It is not possible to proceed with the Taylor expansion of $I(a)$ to still 
higher orders of $a$. This is possibly due to the fact that the next 
contribution in the chiral expansion of $I(a)$ is $\propto a^2{\rm ln}\,a$,
as it is in chiral perturbation theory.


\end{document}